%% file: B2G-21-002_temp.tex
\pdfoutput=1
\documentclass[11pt,twoside,a4paper,cmspaper,final,collab]{cms-tdr}
\def\svnVersion{adac9b8}\def\svnDate{2022/07/07}\def\cmsCernNoTag{CERN-EP-2021-253}\def\cmsCernDate{\today}\def\cmsMessage{Published in Physical Review D as \href{http://dx.doi.org/10.1103/PhysRevD.106.012002}{\doi{10.1103/PhysRevD.106.012002}.}}
\begin{document}\cmsNoteHeader{B2G-21-002}

\newlength\cmsTabSkip\setlength{\cmsTabSkip}{1ex}
\providecommand{\cmsTable}[1]{\resizebox{\textwidth}{!}{#1}}
\newlength\triFigWidth
\ifthenelse{\boolean{cms@external}}{\setlength\triFigWidth{0.8\columnwidth}}{\setlength\triFigWidth{0.32\textwidth}}
\newlength\cmsFigWidth
\ifthenelse{\boolean{cms@external}}{\setlength\cmsFigWidth{0.9\columnwidth}}{\setlength\cmsFigWidth{0.7\textwidth}}
\ifthenelse{\boolean{cms@external}}{\providecommand{\shapeFigLeft}{Upper\xspace}}{\providecommand{\shapeFigLeft}{Left\xspace}}
\ifthenelse{\boolean{cms@external}}{\providecommand{\shapeFigMiddle}{Middle\xspace}}{\providecommand{\shapeFigMiddle}{Center\xspace}}
\ifthenelse{\boolean{cms@external}}{\providecommand{\shapeFigRight}{lower\xspace}}{\providecommand{\shapeFigRight}{right\xspace}}
\newlength\cmsWkern
\ifthenelse{\boolean{cms@external}}{\setlength\cmsWkern{-0.1667em}}{\setlength\cmsWkern{0pt}}
\newcommand{\dPW}{\ensuremath{\text{deep-}\kern\cmsWkern\PW}\xspace}
\newcommand{\dPWorWH}{\ensuremath{\text{deep-}\kern\cmsWkern\PW\,(\text{deep-}\kern\cmsWkern\PW\PH)}\xspace}
\newcommand{\WKK}{\ensuremath{\PW_{\mathrm{KK}}}\xspace}
\newcommand{\ZKK}{\ensuremath{\PZ_{\mathrm{KK}}}\xspace}
\DeclareRobustCommand{\PR}{{\HepParticle{R}{}{}}\xspace}
\DeclareRobustCommand{\PV}{{\HepParticle{V}{}{}}\xspace}
\newcommand{\MR}{\ensuremath{{m}_{\PR}}\xspace}
\newcommand{\MWKK}{\ensuremath{{m}_{\WKK}}\xspace}
\newcommand{\ST}{\ensuremath{S_{\text{T}}}\xspace}
\newcommand{\msd}{\ensuremath{m_{\mathrm{SD}}}\xspace}
\newcommand{\mJ}{\ensuremath{m_{\text{jet}}}\xspace}
\newcommand{\MJ}{\ensuremath{m_{\text{j}}}\xspace}
\newcommand{\MJmax}{\ensuremath{m^{\text{max}}_{\text{j}}}\xspace}
\newcommand{\MJmin}{\ensuremath{m^{\text{min}}_{\text{j}}}\xspace}
\newcommand{\MJmid}{\ensuremath{m^{\text{mid}}_{\text{j}}}\xspace}
\newcommand{\NJ}{\ensuremath{N_{\text{j}}}\xspace}
\newcommand{\Nj}{\NJ}
\newcommand{\NB}{\ensuremath{N_{\PQb}}\xspace}
\newcommand{\ptX}[1]{\ensuremath{\pt^{#1}}\xspace}
\newcommand{\ptj}{\ensuremath{\ptX{\text{j}}}\xspace}
\newcommand{\MJlv}{\ensuremath{m_{\mathrm{j}\Pell\PGn}}\xspace}
\newcommand{\Mjjlv}{\ensuremath{m_{\mathrm{jj}\Pell\PGn}}\xspace}
\newcommand{\Mjj}{\ensuremath{m_{\mathrm{jj}}}\xspace}
\newcommand{\Mjjj}{\ensuremath{m_{\mathrm{jjj}}}\xspace}
\newcommand{\MJJ}{\Mjj}
\newcommand{\MJJJ}{\Mjjj}
\newcommand{\Rlqq}{\ensuremath{\PR^{\Pell\PQq\PQq}}\xspace}
\newcommand{\Rthreeq}{\ensuremath{\PR^{3\PQq}}\xspace}
\newcommand{\Rfourq}{\ensuremath{\PR^{4\PQq}}\xspace}
\newcommand{\PSonel}{\ensuremath{\text{PS}_{1\Pell}}\xspace}
\newcommand{\qg}{\ensuremath{\PQq/\Pg}\xspace}
\newcommand{\Wjets}{\ensuremath{\PW\text{+jets}}\xspace}
\newcommand{\Zjets}{\ensuremath{\PZ\text{+jets}}\xspace}

\cmsNoteHeader{B2G-21-002}
\title{Search for resonances decaying to three \texorpdfstring{\PW}{W} bosons in the hadronic final state in proton-proton collisions at \texorpdfstring{$\sqrt{s} = 13\TeV$}{sqrt(s) = 13 TeV}}

\date{\today}

\abstract{
A search for Kaluza--Klein excited vector boson resonances, \WKK, decaying in cascade to three \PW bosons via a scalar radion \PR, $\WKK \to \PW\PR \to \PW\PW\PW$, in a final state containing two or three massive jets is presented.
The search is performed with $\sqrt{s} = 13\TeV$ proton-proton collision data collected by the CMS experiment at the CERN LHC during 2016--2018, corresponding to an integrated luminosity of 138\fbinv.
Two final states are simultaneously probed, one where the two \PW bosons produced by the \PR decay are reconstructed as separate, large-radius, massive jets, and one where they are merged into a single large-radius jet.
The observed data are in agreement with the standard model expectations.
Limits are set on the product of the \WKK resonance cross section and branching fraction to three \PW bosons in an extended warped extra-dimensional model and are the first of their kind at the LHC.
}

\hypersetup{
pdfauthor={CMS Collaboration},
pdftitle={Search for resonances decaying to three W bosons in the hadronic final state in proton-proton collisions at sqrt(s) = 13 TeV},
pdfsubject={CMS},
pdfkeywords={CMS, vector boson, merged jet}
}

\maketitle
\section{Introduction}
The search for physics beyond the standard model (SM) is one of the most important elements of the research program at the CERN LHC.
Direct searches performed at the LHC have not yet found any compelling evidence for such new physics.
However, novel ideas and recently developed techniques expand the potential for discovery.
For example, in the CMS Collaboration, deep machine learning techniques for tagging Lorentz-boosted resonances decaying hadronically~\cite{JME-18-002} have been developed and exploited extensively for both searches beyond the SM and measurements of the properties of the Higgs boson (\PH)~\cite{Sirunyan:2019qia}.
New physics scenarios involving yet-unprobed signatures of resonant triboson final states through a two-step cascade decay of heavy resonances
in extended warped extra-dimensional models~\cite{Agashe:2016rle,Agashe:2016kfr,Agashe:2017wss,Agashe:2018leo,Kuang:2015gka,Ren:2014bza} have recently been proposed.
These models provide an attractive extension of the SM, which addresses the Planck-electroweak scale difference and flavor hierarchy problems simultaneously.
The theory model probed assumes a Randall--Sundrum scenario with an extended bulk consisting of two extra branes other than the one on which the SM resides~\cite{Agashe:2016rle}.
Only the electroweak gauge fields can propagate into the extended bulk.
The size of the extra dimension is stabilized with a mechanism introducing a potential with a modulus field~\cite{Goldberger_1999}, resulting in a bulk scalar boson, the radion, for each additional brane.
Such extended models can also incorporate heavy resonances that have enhanced decays into triboson final states as compared with direct decays into dibosons and top quark-antiquark pairs.
Thus, a set of new final states emerges with a discovery potential within LHC reach.

In this paper, we report on a search for massive resonances decaying in a cascade into three \PW bosons, through $\WKK \to \PW\PR$ and $\PR \to \PW\PW$,
where \WKK is a Kaluza--Klein (KK) massive excited gauge boson and \PR is a scalar radion.
The analysis is based on proton-proton ($\Pp\Pp$) collision data at $\sqrt{s} = 13\TeV$ collected by the CMS experiment at the LHC during 2016--2018,
corresponding to an integrated luminosity of 138\fbinv.
Since the \WKK excitation has a mass of the order of several \TeVns, the \PW bosons typically have transverse momenta (\pt) of several hundred \GeVns.

\begin{figure}[ht!]\centering
\includegraphics[width=\cmsFigWidth]{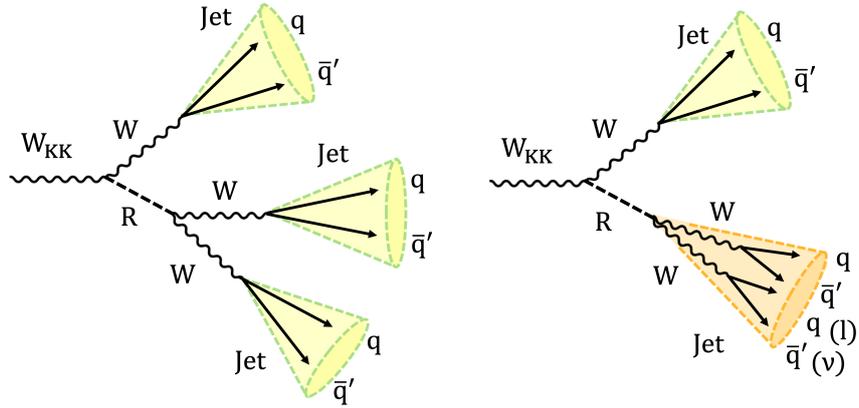}
\caption{Schematic diagrams of the decay of a KK excitation of a \PW boson (\WKK) to the final states considered in this analysis. Additional jets are allowed in the analysis but not considered explicitly.
Left: three individually reconstructed \PW bosons;
right: one individually reconstructed \PW boson and two \PW bosons reconstructed as a single large-radius jet, which is predominant for $\MR \le 0.2\MWKK$.}
\label{fig:diagram}
\end{figure}

In a large fraction of the parameter space ($\MR \lesssim 0.8 \MWKK$), the \PW boson not originating from the radion decay is highly boosted and its decay products are contained in a single large-radius jet.
However, depending on the relative masses of the \WKK and \PR resonances, the two \PW bosons from the \PR decay can either produce two
large-radius jets (``resolved'' case), or one single large-radius jet containing both \PW bosons (``merged'' case).
These two possibilities are illustrated in Fig.~\ref{fig:diagram};
the merged case is predominant when $\MR \le 0.2\,\MWKK$, where \MR and \MWKK are the masses of the \PR and \WKK bosons, respectively.
As a result, the final states considered here require two or three massive jets, predominantly targeting merged and resolved \PR decay topologies, respectively, and no isolated charged leptons.

However, nonisolated leptons are allowed to be present inside the jets formed by merged radion decay products $\PR \to \PW\PW \to \Pell\PGn\PQq\PQq$.
It is also possible to have additional jets in the ``compressed mass'' scenario, $\MR \gtrsim 0.8\,\MWKK$ (depending on the specific value of \MWKK), which can feature
at least one \PW boson with a low boost, whose decay is resolved as two individual small-radius jets.
Such events are not explicitly targeted by this analysis as their production rate is much smaller than the ones of the standard scenarios described above.
This is the first resonance search of this kind in the all-hadronic final state.
In the nonresonant form, as predicted by the SM, the $\PW\PW\PW$ process has recently been observed in final states with at least two charged leptons~\cite{ATLAS:2022xnu,CMS:2020hjs}.

In both cases, merged and resolved, dedicated techniques are applied to exploit the substructure of the \PW boson jets.
For the merged case, apart from the case in which a nonisolated charged lepton overlaps with the hadronically decayed \PW boson, it is also possible that the hadronization products of one or more quarks from the fully hadronic decay $\PR \to \PW\PW \to \PQq\PQq\PQq\PQq$ are not clustered into the same jet.
Events identified as hadronically decaying \PW bosons can also include cases where the decay $\PW \to \PGt\PGn$ is followed by a hadronic decay of the tau lepton.
These effects lead to a complicated jet mass spectrum from the merged radion that requires the design of a hybrid discriminant (``tagger'').
Events with a single isolated charged lepton in the final state are considered in a similar, separate analysis with nonoverlapping event selection, described in Ref.~\cite{CMS-B2G-20-001}.

While the search is by design optimized for a $\PW\PW\PW$ signal, it is also partly sensitive to signals with similar decay topologies.
In particular, heavy resonances decaying into $\PW\PW$, $\PW\PZ$, $\PZ\PZ$, $\PW\PW\PZ$, $\PW\PZ\PZ$, $\PZ\PZ\PZ$, $\PW\PQt$, $\PZ\PQt$, $\PW\PH$, $\PZ\PH$, $\PW\PX$, or $\PZ\PX$,
where \PX denotes an unknown particle with mass above 70\GeV whose decay products can be identified using jet substructure techniques, can be detected through this search.

This paper is organized as follows:
Section~\ref{sec:cms} provides a description of the CMS detector.
Section~\ref{sec:Sample} describes the data sets and simulation samples used in the analysis.
The triggers used for data collection and the event reconstruction are discussed in Section~\ref{sec:Objects}.
The massive jet tagging is described in Section~\ref{sec:Jettagging}.
The event selection and event categorization are presented in Section~\ref{sec:event_selection}.
The jet tagger calibration is described in Section~\ref{sec:Tagger_calibration}.
Section~\ref{sec:bkg_estimation} describes the estimation of the SM background.
Systematic uncertainties are discussed in Section~\ref{sec:uncertainties}.
The results and their interpretation are given in Section~\ref{sec:Results}.
A summary is presented in Section~\ref{sec:sum}.
Tabulated results are provided in the HEPData record for this analysis~\cite{hepdata}.

\section{The CMS detector}
\label{sec:cms}
The central feature of the CMS detector is a superconducting solenoid of 6\unit{m} internal diameter, providing a magnetic field of 3.8\unit{T}.
A silicon pixel and strip tracker, a lead tungstate crystal electromagnetic calorimeter (ECAL), and a brass and scintillator hadron calorimeter (HCAL),
each composed of a barrel and two endcap sections resides within the solenoid volume.
Forward calorimeters extend the coverage provided by the barrel and endcap detectors up to pseudorapidities of $\abs{\eta} = 5$.
Muons are measured in gas-ionization detectors embedded in the steel flux-return yoke outside the solenoid.

Events of interest are selected using a two-tiered trigger system.
The first level, composed of custom hardware processors, uses information from the calorimeters and muon detectors to select events at a rate of around 100\unit{kHz}, making a decision within the fixed period of 4\mus following the beam crossing, allowed by the latency implemented in the readout path~\cite{Sirunyan:2020zal}.
The second level, known as the high-level trigger (HLT), consists of a farm of processors running a version of the full event reconstruction software optimized for fast processing,
and reduces the event rate to around 1\unit{kHz} before data storage~\cite{Khachatryan:2016bia}.
A more detailed description of the CMS detector, together with a definition of the coordinate system and kinematic variables, can be found in Ref.~\cite{Chatrchyan:2008zzk}.

\section{Data samples and event simulation} \label{sec:Sample}

The data samples analyzed in this search correspond to a total integrated luminosity of 138\fbinv.
They were recorded in $\Pp\Pp$ collisions at $\sqrt{s} = 13\TeV$ in the years 2016, 2017, and 2018,
comprising 36.3, 41.5, and 59.7\fbinv, respectively~\cite{Sirunyan:2021qkt,CMS-PAS-LUM-17-004,CMS-PAS-LUM-18-002}.

The signal is simulated at leading order (LO) using the \MGvATNLO 2.4.2 generator~\cite{MGatNLO},
covering a wide range of \WKK and \PR masses (\MWKK from 1.5 to 5.0\TeV, and \MR from 6 up to 90\% of \MWKK),
together with the parameters recommended by the authors of Refs.~\cite{Agashe:2016rle,Agashe:2016kfr,Agashe:2017wss,Agashe:2018leo}, \ie, a KK coupling to the radion and a \PW boson $g_{\text{grav}}=6$, KK gauge couplings $g_{\WKK}=3$ and $g_{\ZKK}=6.708$, and a confinement parameter $\epsilon=0.5$.
The branching fraction for the decay $\WKK \to \PR \PW \to \PW\PW\PW$ can reach values above 50\%.

Top quark pair and single top quark production are modeled at next-to-LO (NLO) using the \POWHEG 2.0 generator~\cite{POWHEG1,POWHEG2,POWHEG3,Alioli:2011as,Alioli:2009je,Frederix:2012dh}.
Events composed uniquely of jets produced through the strong interaction are referred to as quantum chromodynamics (QCD) multijet events.
These processes, along with background from \Wjets and \Zjets production, are simulated at LO with \MGvATNLO, and matched to parton showers with the MLM~\cite{Alwall:2007fs} algorithm.
The other, less important backgrounds, including processes with two or three vector bosons $\PV = \PW, \PZ$  (diboson and triboson production, respectively),
are simulated at NLO with either \POWHEG (\PW{}\PW production) or \MGvATNLO (all others).
The simulation of $\ttbar\PW/\PZ$ events is performed at LO using \MGvATNLO.

All background and signal samples for the 2016 data-taking conditions are generated with the NNPDF3.0 NLO or LO parton distribution functions (PDFs)~\cite{nnpdf}, with the order matching that in the matrix element calculations.
To model processes in the 2017 and 2018 data sets, the NNPDF3.1 next-to-next-to-LO PDFs~\cite{Ball:2017nwa} are used for all samples.
Parton showering, fragmentation, and hadronization for all samples are performed using \PYTHIA 8.230~\cite{Sjostrand:2014zea} with the underlying event tune CUETP8M1~\cite{Khachatryan:2015pea}
for the 2016 analysis, and CP5~\cite{Sirunyan:2019dfx} for the 2017 and 2018 analyses.
The CMS detector response is modeled using the \GEANTfour package~\cite{geant4, geant4_2}.
A tag-and-probe procedure~\cite{tagandprobe} is used to derive corrections for data-to-simulation differences in reconstruction and selection efficiencies.
The simulated events include additional $\Pp\Pp$ interactions in the same and neighboring bunch crossings, referred to as pileup (PU).
The simulated events are weighted so the PU vertex distribution matches the one from the data.

\section{Event reconstruction} \label{sec:Objects}

The candidate vertex with the largest value of summed physics-object $\pt^2$ is taken to be the primary $\Pp\Pp$ interaction vertex.
The physics objects used for this determination are the jets, clustered using the anti-\kt jet finding algorithm~\cite{Cacciari:2008gp,Cacciari:2011ma} with the tracks assigned to candidate vertices as inputs,
and the associated missing transverse momentum (\ptvecmiss), taken as the negative vector sum of the \pt of those jets.

A particle-flow (PF) algorithm~\cite{CMS-PRF-14-001} aims to reconstruct and identify each interacting particle in an event,
with an optimized combination of information from the various elements of the CMS detector.
The energy of electrons is determined from a combination of the track momentum at the primary interaction vertex, the corresponding ECAL cluster energy, and the energy sum of all bremsstrahlung photons attached to the track.
The energy of muons is obtained from the curvature of the corresponding track.
The energy of charged hadrons is determined from a combination of their momentum measured in the tracker and the matching ECAL and HCAL energy deposits,
corrected for the response function of the calorimeters to hadronic showers.
Finally, the energy of neutral hadrons is obtained from the corresponding corrected ECAL and HCAL energies.

For each event, hadronic jets are clustered from these reconstructed particles using the infrared and collinear safe anti-\kt algorithm~\cite{Cacciari:2008gp, Cacciari:2011ma}.
The clustering algorithm is run twice over the same inputs, once with a distance parameter of 0.4 (AK4 jets) and once with 0.8 (AK8 jets).
Jet momentum is determined as the vectorial sum of all particle momenta in the jet, and is found from simulation to be, on average, within 5 to 10\% of the true momentum over the entire \pt spectrum and detector acceptance.

Pileup interactions can contribute additional tracks and calorimetric energy depositions to the jet momentum.
The pileup per particle identification algorithm (PUPPI)~\cite{Sirunyan:2020foa,Bertolini:2014bba} is used to mitigate the effect of PU at the reconstructed particle level.
Using this algorithm, the momenta of charged and neutral particles are rescaled.
Jet energy corrections are derived from simulation to bring the measured response of jets to that of particle-level jets on average.
In situ measurements of the momentum balance in dijet, $\text{photon}$+jet, $\PZ$+jet, and multijet events are used to account for any residual differences in the jet energy scale between data and simulation~\cite{Khachatryan:2016kdb}.
The jet energy resolution amounts typically to 15--20\% at 30\GeV, 10\% at 100\GeV, and 5\% at 1\TeV~\cite{Khachatryan:2016kdb}.
Additional selection criteria are applied to each jet to remove jets potentially dominated by anomalous contributions from various subdetector components or reconstruction failures~\cite{CMS-PAS-JME-16-003}.

Jets originating from the hadronization of \PQb quarks are identified using a deep neural network algorithm (\textsc{DeepCSV}) that takes as input: tracks displaced from the primary interaction vertex,
identified secondary vertices, jet kinematic variables, and information related to the presence of soft leptons in the jet~\cite{Sirunyan:2017ezt}.
Working points (WPs) are used that yield either a 1\% (medium WP) or a 10\% (loose WP) probability of misidentifying a light flavor quark or a gluon ($\PQu\PQd\PQs\Pg$) AK4 jet with $\pt > 30\GeV$ as a \PQb quark jet.
The corresponding average efficiencies for the identification of the hadronization products of a bottom quark as a \PQb quark jet are about 70 and 85\%, respectively.

The vector \ptvecmiss is computed as the negative vector sum of the transverse momenta of all the PF candidates in an event, and its magnitude is denoted as \ptmiss~\cite{Sirunyan:2019kia}.
The \ptvecmiss is modified to account for corrections to the energy scale of the reconstructed jets in the event.
Anomalous high-\ptmiss events can be due to a variety of reconstruction failures, detector malfunctions, or noncollision backgrounds.
Such events are rejected by event filters that are designed to identify more than 85--90\% of the spurious high-\ptmiss events with a mistagging rate of less than 0.1\%~\cite{Sirunyan:2019kia}.

Hadronic decays of $\PW/\PZ$ bosons are identified with the groomed jet mass (\MJ) and a novel deep learning algorithm with the PF candidates and secondary vertices as inputs~\cite{JME-18-002}.
The groomed jet mass is calculated after applying a modified mass-drop algorithm~\cite{Dasgupta:2013ihk,Butterworth:2008iy} to AK8 jets, with parameters $\beta = 0$, $z_\text{cut} = 0.1$, and $R_0 = 0.8$. This algorithm is also known as the \emph{soft-drop} algorithm~\cite{Larkoski:2014wba}.
The variables are calibrated in a top quark-antiquark sample enriched in hadronically decaying \PW bosons~\cite{Khachatryan:2014vla}.
Further details on the calibration method used for this analysis are given in Section~\ref{sec:Tagger_calibration}.

Muon (\PGm) and electron (\Pe) candidates are reconstructed in order to veto events containing such energetic leptons.
Muon candidates are required to be within the geometrical acceptance of the muon detectors ($\abs{\eta} < 2.4$) and are reconstructed by combining the information from the silicon tracker and the muon chambers~\cite{Sirunyan:2018fpa}.
These candidates are required to satisfy a set of quality criteria based on the number of hits measured in the silicon tracker and in the muon system,
the properties of the fitted muon track, and the transverse and longitudinal impact parameters of the track with respect to the primary vertex of the event.
Electron candidates within ${\abs{\eta} < 2.5}$ are reconstructed using an algorithm that associates fitted tracks in the silicon tracker with electromagnetic energy clusters in the ECAL~\cite{Khachatryan:2015hwa}.
To reduce the misidentification rate, these candidates are required to satisfy identification criteria based on the shower shape of the energy deposit,
the matching of the electron track to the ECAL energy cluster, the relative amount of energy deposited in the HCAL detector, and the consistency of the electron track with the primary vertex.
Because of nonoptimal reconstruction performance, electron candidates in the transition region between the ECAL barrel and endcaps, $1.44 < \abs{\eta} < 1.57$, are discarded.
Electron candidates identified as coming from photon conversions in the detector are also rejected.
Identified muons and electrons are required to be isolated from hadronic activity in the event.
The isolation sum is defined by summing the \pt of all the PF candidates in a cone of radius ${\Delta R = \sqrt{\smash[b]{(\Delta\eta)^{2}+(\Delta\phi)^{2}}} = 0.4\,(0.3)}$ around the muon (electron) track,
and is corrected for the contribution of neutral particles from PU interactions~\cite{Sirunyan:2018fpa,Khachatryan:2015hwa}.

\section{Massive jet tagging} \label{sec:Jettagging}

The signal event signatures include two types of massive jets ($\MJ>60\GeV$) originating from the merged decay products of either \PW or \PR bosons.
We consider three main cases for the merged \PR boson decay, designated and defined as follows:
\begin{itemize}
\item \Rfourq, where the two daughter \PW bosons decay hadronically ($\PR \to \PW\PW \to \Pq\Pq\Pq\Pq$) with all four final-state quarks contained in the reconstructed jet
\item \Rthreeq, similar to the former but with one quark leaking outside of the jet cone, producing a three-prong jet
\item \Rlqq, where one of the two daughter \PW bosons decays leptonically ($\PR \to \PW\PW \to \Pell\PGn \Pq\Pq$), resulting in a jet containing an energetic, charged, nonisolated lepton
\end{itemize}
All these types of \PR candidate jets are reconstructed as AK8 jets.

Both \PW and \PR boson candidates are tagged using the \textsc{DeepAK8} jet classification framework~\cite{JME-18-002}.
This modular tagging framework has been designed by the CMS Collaboration to identify hadronically decaying top quarks as well as \PW, \PZ, and Higgs bosons.
The algorithm uses machine learning techniques based on PF candidates, secondary vertices, and other inputs to classify the AK8 jets into 17 categories.
These categories include jets arising from $\PW\to \Pq\Pq$, $\PZ\to \Pq\Pq$, $\PQt\to \PQb\Pq\Pq$, $\PH\to 4\Pq$, and gluon or light-quark decay.
To remove a potential mass dependence from the classifier output, a generative adversarial neural network is used to create ``mass-decorrelated'' outputs.
The final output is a set of 17 ``raw scores'' per jet, where each one gives the likelihood of the jet originating from a particular decay.
Discriminants have been developed by summing these raw scores and taking appropriate ratios to select particular types of jets, while rejecting others.

Two particular discriminants are used for this analysis.
The first, ``\dPW'', aims to identify \PW boson candidates through the $\PW \to \PQq\PQq$ and QCD multijet raw scores, selecting and rejecting compatible jets, respectively.
The second, ``$\dPW\PH$'', is used to identify merged \PR boson candidates of types $\PR^{\Pell\Pq\Pq}$, $\PR^{3\Pq}$, and $\PR^{4\Pq}$.
This is achieved by making use of the $\PW\to \Pq\Pq$ and $\PH\to 4\Pq$ raw scores, which select radion-like jet types while rejecting QCD multijet candidates.
The corresponding formulae are as follows:
\begin{linenomath}
\begin{equation}
\dPW = \frac{\text{raw score}(\PW \to \PQq\PQq)}{\text{raw score}(\PW \to \PQq\PQq) + \text{raw score(QCD)}},
\end{equation}
\end{linenomath}
used for tagging \PW boson candidates with mass \MJ in the range 60--100\GeV, and
\begin{linenomath}

\begin{equation}
    \dPW\PH = \frac{\text{r.s.}(\PW \to \PQq\PQq) + \text{r.s.}(\PH \to 4\Pq)}{\text{r.s.}(\PW \to \PQq\PQq) + \text{r.s.}(\PH \to 4\Pq) + \text{r.s.(QCD)}},
\end{equation}
\end{linenomath}
where ``r.s.'' denotes the raw score,
used for tagging radion candidates with mass $\MJ > 100\GeV$.

For both taggers, the mass-decorrelated version is used to avoid distorting the mass distribution (mass sculpting) and to retain the sensitivity to radions with mass greater than those of the \PW and Higgs bosons.
The tagger discriminant distributions are presented in Fig.~\ref{fig:PreselectionPlots} (lower row) using a loose selection that will be defined in Section~\ref{subsec:Preselection_and_SR}.

\section{Event selection} \label{sec:event_selection}

\subsection{Trigger} \label{subsec:Trigger}

The analysis uses events that are selected by a range of different HLT paths.
One set of paths requires \HT, the scalar sum of the \pt of all AK4 jets in the event, to be greater than 800, 900, or 1050\GeV,
depending on the data collection year.
In addition, events with $\HT > 650\GeV$ and a pair of jets with invariant mass above 900\GeV and a pseudorapidity separation $\abs{\Delta\eta} < 1.5$ are also selected for the 2016 data set.
A different set of paths selects events where the \pt of the leading AK8 jet is greater than 500\GeV, or the \pt is greater than 360\GeV and the ``trimmed mass'' of an AK8 jet is above 30\GeV.
The jet trimmed mass is obtained after removing remnants of soft radiation with the jet trimming technique~\cite{Krohn:2009th},
using a subjet size parameter of 0.3 and a subjet-to-AK8 jet \pt fraction of 0.1.
The trigger selection efficiency is measured to be greater than 99\% for events with $\HT > 1.1\TeV$, using an independent sample of data events collected with a single-muon trigger.

\subsection{Preselection and signal region} \label{subsec:Preselection_and_SR}

Events are selected in two stages; the first, ``preselection'', is initially applied to explore kinematic features of the signal compared to the SM background.
A tighter selection, the signal region (SR) selection, is then applied to further improve the background rejection.
The final analysis uses the SR events, while the preselected events are used to calibrate and validate the \textsc{DeepAK8} discriminants \dPW and $\dPW\PH$.
In the following, we simply use the term ``jets'' to indicate massive AK8 jets if not stated differently.

The following kinematic conditions define the preselection:
\begin{itemize}
\item jet $\pt^{\text{j}}>200\GeV$;
\item number of jets, \NJ, exactly 2 or 3;
\item highest \pt jet $\pt^{\text{j1}} > 400\GeV$;
\item mass of the two highest \pt jets $m_{\text{j1,j2}}>40\GeV$;
\item no isolated lepton ($N_{\Pell}=0$) with $\pt^{\Pell}>20\,(35)\GeV$ and $\abs{\eta^{\Pell}} < 2.4\,(2.5)$ for \PGm (\Pe).
\end{itemize}

\begin{figure*}[htbp]\centering
\includegraphics[width=0.42\linewidth]{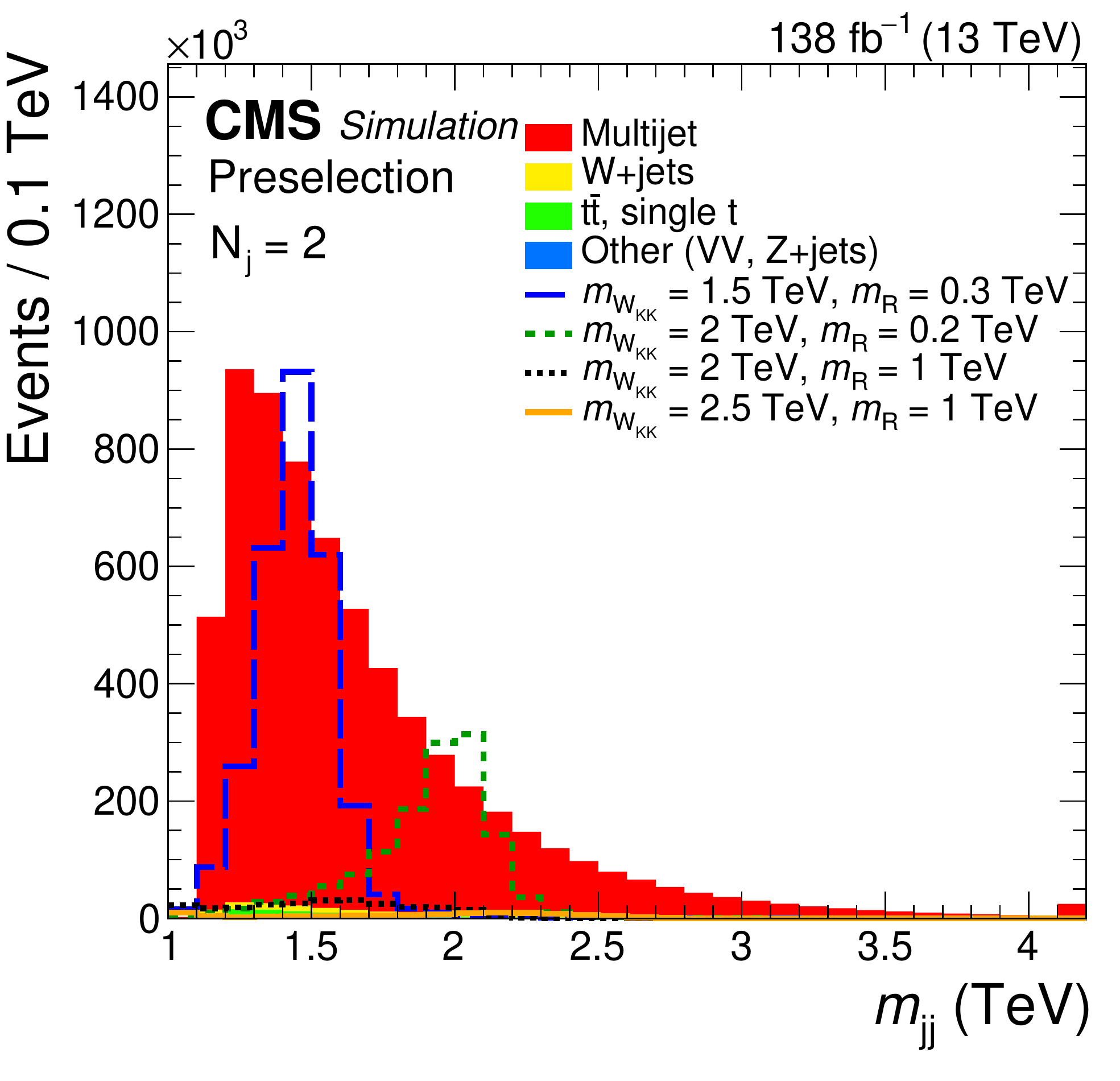}
\includegraphics[width=0.42\linewidth]{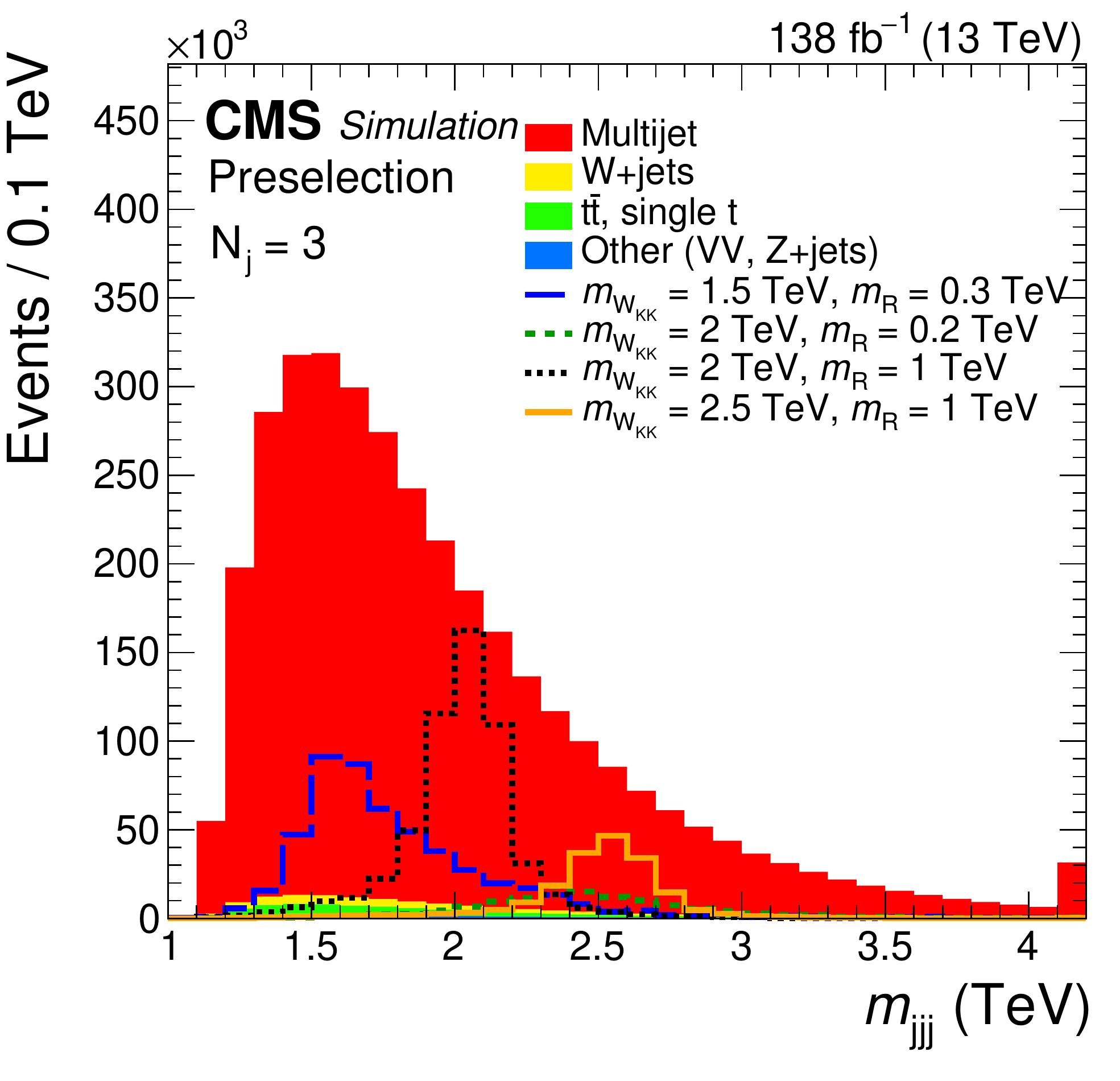}\\
\includegraphics[width=0.42\linewidth]{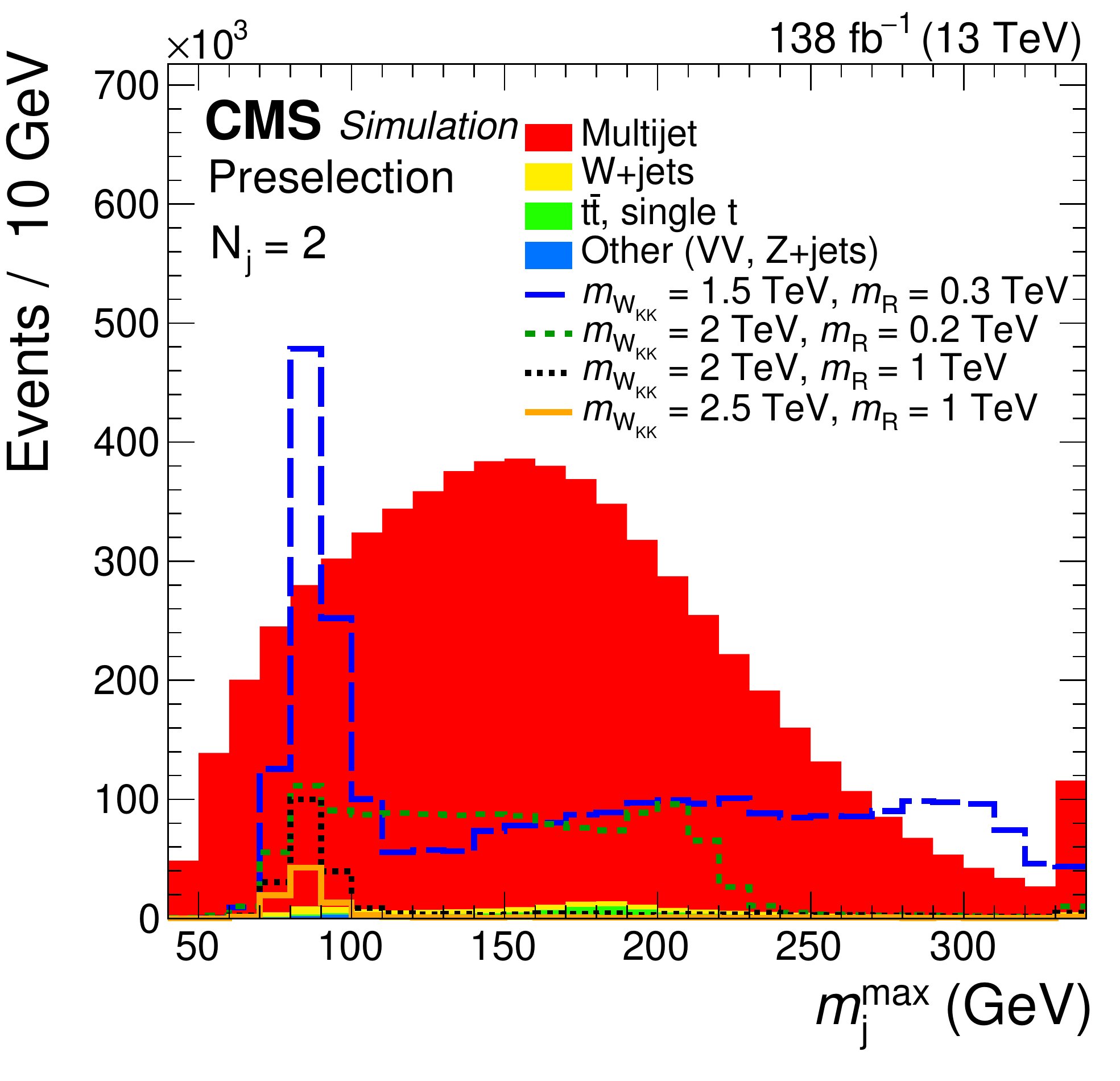}
\includegraphics[width=0.42\linewidth]{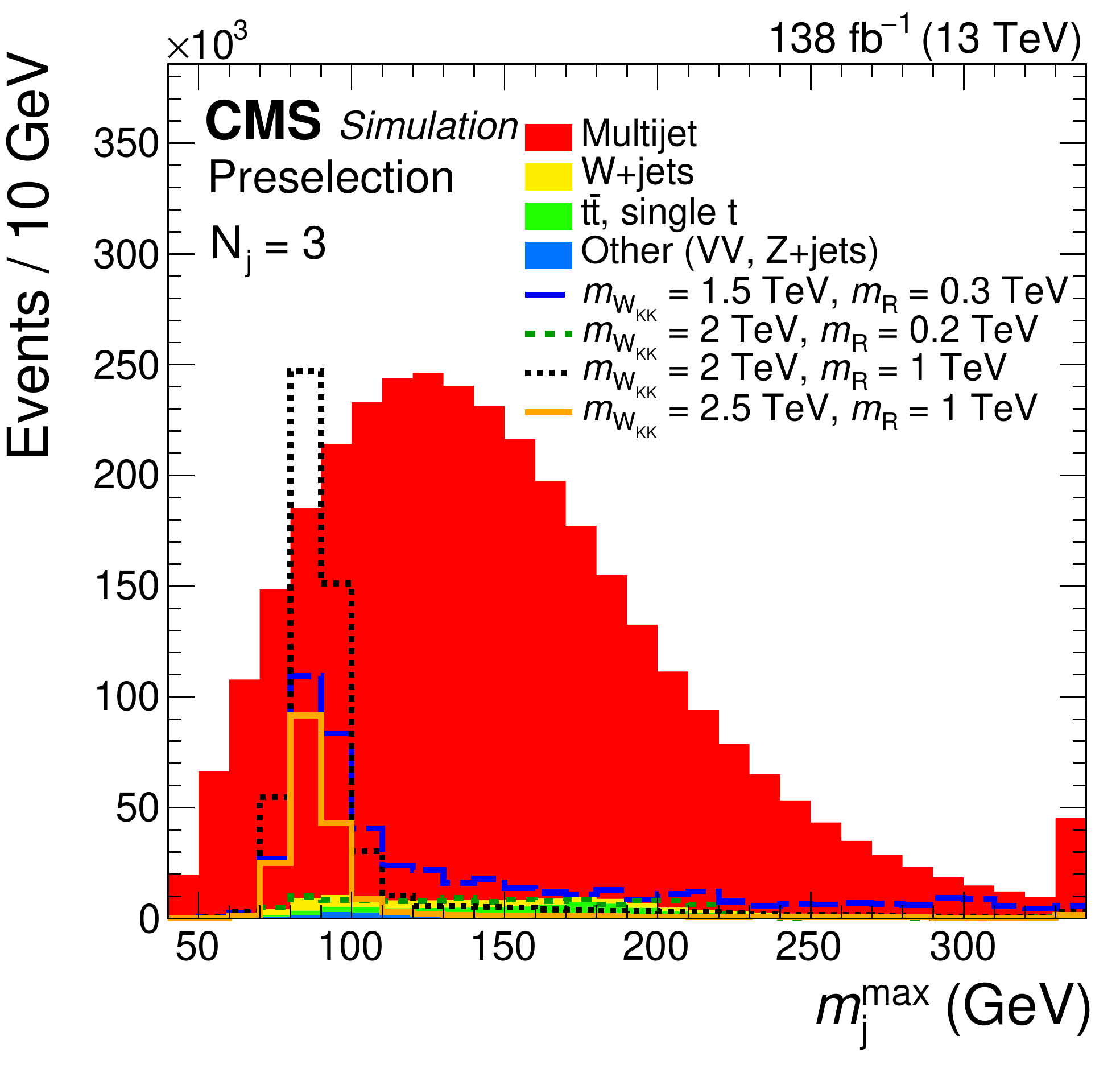}\\
\includegraphics[width=0.42\linewidth]{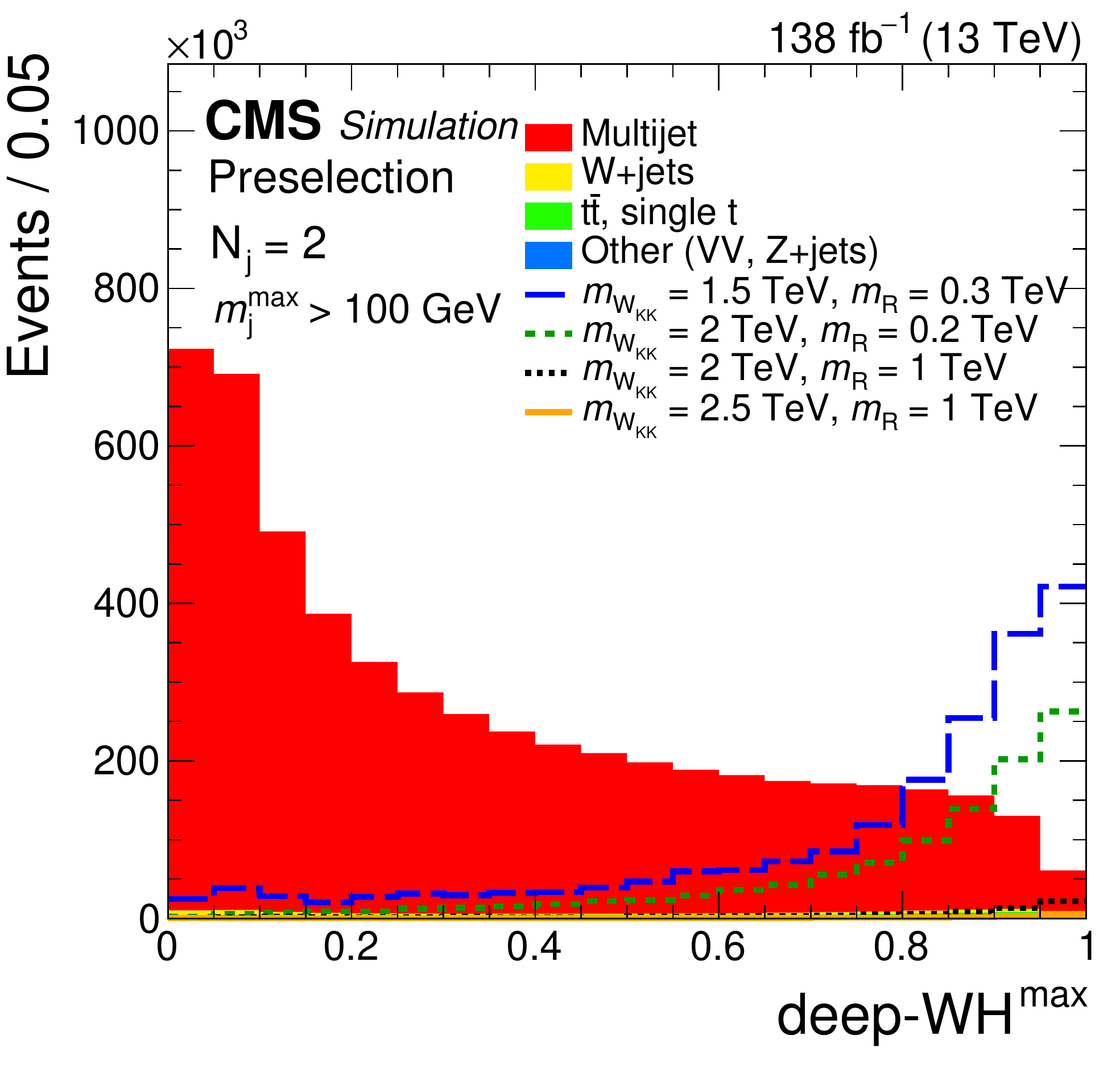}
\includegraphics[width=0.42\linewidth]{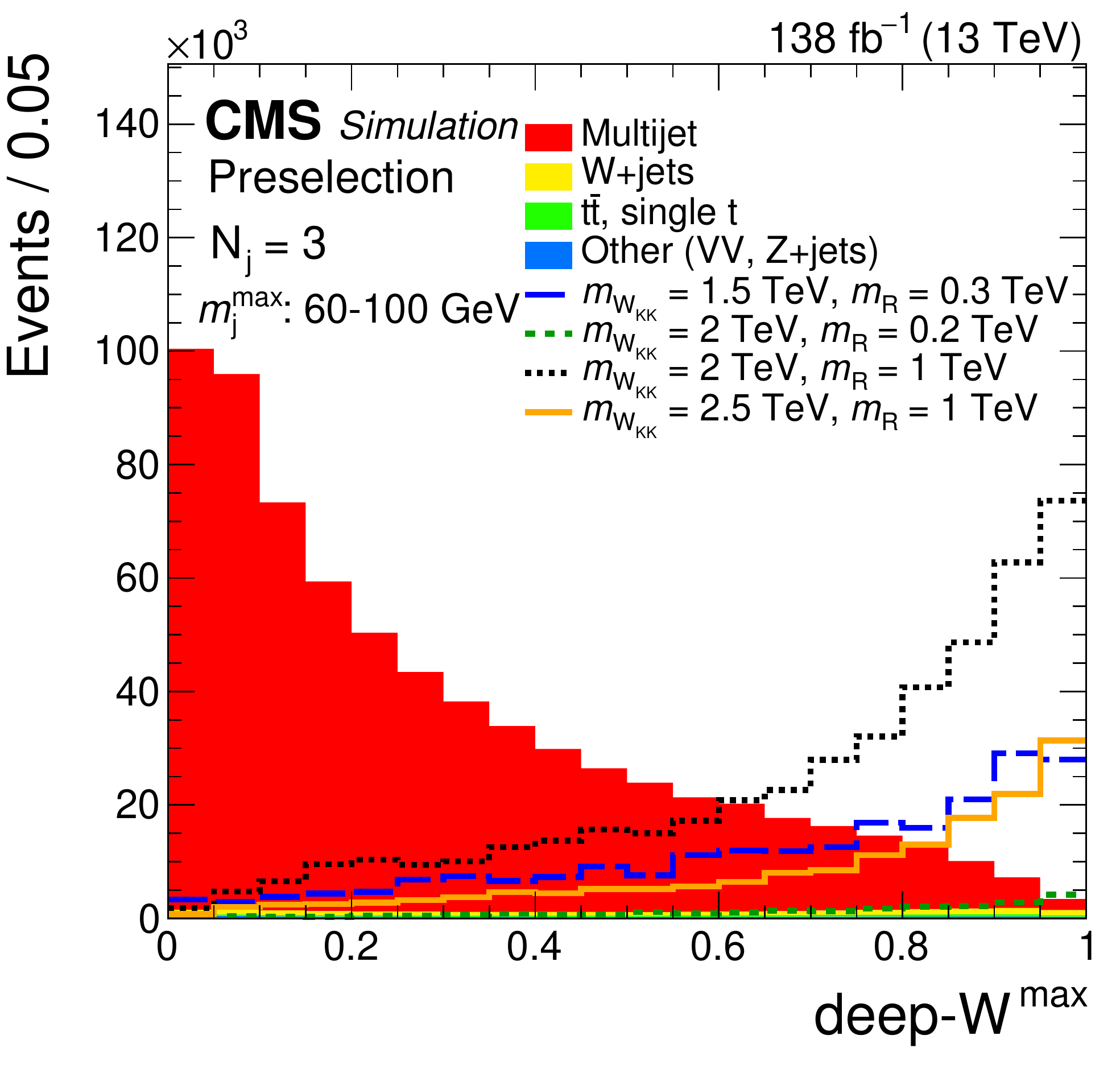}
\caption{Variables discriminating between signal and background in simulation.
Left column, upper to lower rows: the distributions of \Mjj, \MJmax, and $\dPW\PH$ (for highest mass jet with $\MJmax>100\GeV$) for preselected events with $\NJ=2$.
Right column, upper to lower rows: the distributions of \Mjjj, \MJmax, and \dPW (for highest mass jet with $60<\MJmax<100\GeV$) for preselected events with $\NJ=3$.
The signal processes are scaled to 500 times their theoretical cross sections for visibility.}
\label{fig:PreselectionPlots}
\end{figure*}

The triboson signal is expected to show a peak in the distribution of the invariant mass of the jets, \Mjj in dijet events and \Mjjj in trijet events.
These distribution are used for the statistical analysis.
Figure~\ref{fig:PreselectionPlots} (upper row) shows the \Mjj (\Mjjj) spectra for signal and background after preselection.
The signal processes are scaled to 500 times their theoretical cross sections.

To define the SR selection, we add the following conditions to the preselection criteria.
In the case of $\NJ=2$ events, the higher and lower jet masses are designated as \MJmax and \MJmin, respectively. The higher-mass jet is taken to be the radion candidate, and the lower-mass jet to be the \PW boson candidate.
Therefore, we require $\MJmax>70\GeV$ and $70<\MJmin<100\GeV$.
In the case of events with $\NJ=3$, \MJmax and \MJmin are defined as above, with \MJmid designating the jet intermediate in mass. The two highest mass jets are considered as \PW boson candidates, and we demand $70<(\MJmid, \MJmax)<100\GeV$.
The lowest mass jet is required to have mass $\MJmin<100\GeV$. This jet can correspond to either a merged \PW boson ($60<\MJmin<100\GeV$) or to a single quark originating from a \PW boson decay ($\MJmin<60\GeV$).
Therefore, we allow at most one of the three \PW bosons to be resolved into a pair of low-mass jets ($\MJ<60\GeV$) with exactly one of the two daughter-quark jets required to have \pt above the 200\GeV threshold.
Figure~\ref{fig:PreselectionPlots} (middle row) shows the \MJmax distributions.

Jets in the mass range 60--100 ($>$100)\GeV, as \PW boson (radion) candidates, are further selected using the \dPWorWH discriminant.
Figure~\ref{fig:PreselectionPlots} (lower row) presents the \dPWorWH distributions for the highest mass jets after preselection.
The conditions $\dPW>0.8$ and $\dPW\PH>0.8$ are required for events with two massive jets, while the less stringent requirement of at least two massive jets with $\dPW>0.6$ is imposed for events with three jets.

In order to select Lorentz-boosted final states, we additionally require that \ST, the scalar sum of the transverse momenta of the selected jets and the \ptmiss, is greater than 1.3\TeV.
The \ptmiss in the \ST sum enhances signal separation for the cases where a hadronic \PGt lepton decay is present, or where the decay products from a merged radion decay include a nonisolated lepton, since in these cases the \ptmiss arises from the undetected neutrino(s).
To suppress \ttbar background, events are vetoed that contain a \PQb-tagged AK4 jet not overlapping with any AK8 jet ($\Delta R>0.8$).
The \textsc{DeepCSV} discriminant at the medium working point~\cite{Sirunyan:2017ezt} is used for this veto.
As the signal region explored corresponds to $\MWKK\ge 1.5\TeV$, we also impose the condition $\Mjj(\Mjjj)>1.1\TeV$ to probe only the high-mass region, although this condition has minimal impact on top of the \ST and \HT constraints.
While the selection requirements do not explicitly target the case where the lowest \pt \PW boson is resolved into two single-quark jets, some of these events are accepted if only one of the two single-quark jets has $\pt>200\GeV$.

The SR selection criteria, which are applied on top of preselection, can be summarized as:
\begin{itemize}
\item Number of additional \PQb-tagged jets (nonoverlapping with the AK8 jets) $N_{\PQb} = 0$ (medium WP)
\item Sum of \ptmiss and the \pt of the selected jets: $\ST > 1.3\TeV$
\item Dijet (trijet) invariant mass $\MJJ (\MJJJ) > 1.1\TeV$ for $\NJ = 2$ (3)
\item For $\NJ$ = 2: $\MJmax > 70\GeV$, $70 < \MJmin < 100\GeV$, with $\dPW (\PW\PH) > 0.8$ for $70 < \MJ < 100\GeV$ ($\MJ > 100\GeV$)
\item For $\NJ$ = 3: $70 < (\MJmax$, $\MJmid) < 100\GeV$ and $\MJmin < 100\GeV$, with $\dPW > 0.6$ (0.8) for three (two) massive jets
\end{itemize}

\subsection{Signal region definition}\label{subsec:SR_Categorization}

\begin{figure*}[ht!] \centering
    \includegraphics[width=0.9\textwidth]{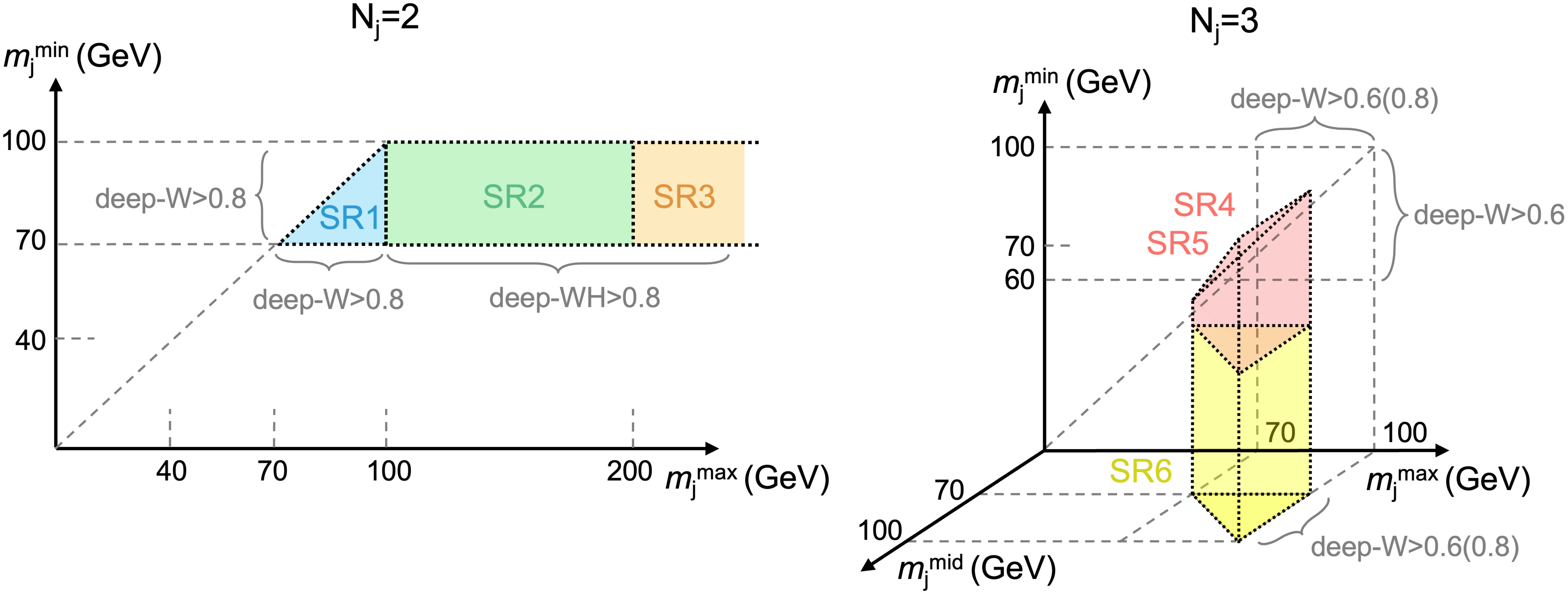}
    \caption{Schematic of the 2D jet mass regions for two-jet events (left) and 3D jet mass regions for three-jet events (right), indicating the location of the six independent signal regions SR1--6, indicated by the colored areas.
    The SR4 and SR5 differ by the requirement of exactly three and two \PW-tagged jets, respectively.
    The jet tagging discriminants used in the event selection  are also shown for each of the mass-ordered jets. The values in parentheses indicate that, depending on the SR, different selection requirements are employed.}
    \label{fig:SR_cartoon}
\end{figure*}

Six different SRs are defined in the following and are summarized in Table~\ref{tab:SRs}.
In addition, Fig.~\ref{fig:SR_cartoon} illustrates these SRs in two-dimensional (2D) and 3D diagrams of the jet mass.

\begin{table*}[thb]
    \centering
    \topcaption{Summary of the selection requirements for each of the signal regions.}
    \label{tab:SRs}
    \cmsTable{
    \begin{scotch}{cccccc}
    Region & \NJ & \MJmax (\GeVns{}) & \MJmid (\GeVns{}) & \MJmin (\GeVns{}) & Jet tagging conditions \\ \hline
    SR1 & 2 & 70--100  &   \NA   & 70--100 & Both with $\dPW > 0.8$  \\
    SR2 & 2 & 100--200 &   \NA   & 70--100 & Higher with $\dPW\PH > 0.8$, lower with $\dPW > 0.8$  \\
    SR3 & 2 & $>$200   &   \NA   & 70--100 & Higher with $\dPW\PH > 0.8$, lower with $\dPW > 0.8$  \\ [\cmsTabSkip]
    SR4 & 3 & 70--100  & 70--100 & 60--100 & All three with $\dPW > 0.6$  \\
    SR5 & 3 & 70--100  & 70--100 & 60--100 & Exactly two with $\dPW > 0.6$ \\
    SR6 & 3 & 70--100  & 70--100 &  0--60  & Two highest with $\dPW > 0.8$ \\
    \end{scotch}
    }
\end{table*}

The SR events with $\Nj = 2$ are split into three samples based on the value of \MJmax: SR1, SR2, and SR3 correspond to \MJmax values of 70--100, 100--200, and $>$200\GeV, respectively.
This categorization serves as a binning over the unknown radion mass.
As Fig.~\ref{fig:PreselectionPlots} (middle row) illustrates, the merged radion jet mass has a broad distribution populating the \MJmax range of 70\GeV to \MR.
Signal events in SR2 and SR3 (\ie, with $\MJmax > 100\GeV$) generally contain a merged radion jet (\Rlqq, \Rthreeq, \Rfourq), and the $\dPW\PH$ discriminant separates these jets from the SM background.

Events in SR1 have both jets in the 70--100\GeV mass window.
The merged radion jet lies in SR1 either for cases where the higher-mass jet is in the \Rlqq category and the neutrino acquires most of the parent \PW boson momentum,
or when the higher-mass jet is a \PW boson jet (when the decay products of $\PR\to \PW\PW$ receive imbalanced Lorentz boosts and the softer \PW boson is not merged).
Resolved-radion events, \ie, events where the radion is reconstructed as two \PW boson jets, can lie in SR1 if the softest hadronically decaying \PW boson (typically the one produced promptly from the \WKK decay) is not reconstructed as a single jet and therefore not selected as a candidate jet.
In addition, SR1 is sensitive to any diboson resonant signal that might be present.
Any jet of SR1--3 with a mass in the range 70--100\GeV is required to satisfy the $\dPW > 0.8$ requirement to be tagged as a \PW (or \Rlqq) boson candidate.

The SR events with $\Nj = 3$ are split into three samples SR4--6 as follows.
In the case of $\MJmin > 60\GeV$, we demand all three jets to be \PW-tagged satisfying the condition $\dPW > 0.6$, which defines the SR4 region; events with exactly two \PW-tagged jets are placed in SR5.
Events with $\MJmin < 60\GeV$ and the other two massive jets satisfying $70 < (\MJmid, \MJmax) < 100\GeV$ and $\dPW > 0.8$ constitute SR6.
These three regions are sensitive only to the resolved-radion signal, with SR4 being the most sensitive among them, as it demands three \PW-tagged jets.

The six different regions provide complementary sensitivity to different regions of the \MWKK-\MR plane.
Signal scenarios with radion masses producing jets in the mass range 100--200\GeV are predominantly probed in SR2;
for \MR in the range 200--300\GeV, the signal events predominantly lie in SR2 and SR3;
while for $\MR > 300\GeV$ (if the radion remains merged), SR3 provides most of the sensitivity.

\section{Calibration of the \textsc{DeepAK8} tagger} \label{sec:Tagger_calibration}

The \dPWorWH discriminants are not fully reproduced in simulation, especially at low and high scores.
In this section we describes the calibration procedure followed to correct the \dPWorWH spectra for each type of jet in two bins of \ptj and \MJ.
All types of jets involved in this procedure are illustrated in Table~\ref{tab:Matching}.
The correction is quantified using scale factors (SFs), which are applied to all simulated events (signal and background).

Events in the preselected sample are dominated by QCD multijet background (99\%). In SR1--6, QCD multijet events make up 50--75\% of the expected background.
The rest of the events are from \ttbar and single \PQt quark processes (10--25\%), \Wjets processes (10--20\%), and other processes (\eg, $\PW\PW$, $\PW\PZ$, $\PQt\PAQt\PW/\PZ$, or tribosons, making up less than 15\%).
Therefore, massive jets ($\MJ > 60\GeV$) selected in the SRs are predominantly a mixture of different jet categories that we define as follows:
\begin{itemize}
\item{hadronically decaying \PW bosons producing merged \PW boson jets}
\item{light quarks or gluons (\qg), with radiation or fragmentation, which are reconstructed as massive \qg jets}
\item{three types of jets from hadronically decaying \PQt quarks, $\PQt \to \PQb\PW \to \PQb\PQq\PQq$:}
\begin{itemize}
\item{jets including the \PQb quark and only one of the quarks from the \PW boson decay, designated ``$\PQt^2$''}
\item{jets including the \PQb quark and both of the quarks from the \PW boson decay, designated ``$\PQt^3$''}
\item{same as ``$\PQt^3$'', but requiring an additional energetic quark or gluon inside the jet cone to define a four-prong category, designated ``$\PQt^4$''}
\end{itemize}
\end{itemize}
For the $\PQt^4$ category, the additional \qg inside the jet cone needs to have $\pt>50\GeV$.
By considering $\PQt^3$ and $\PQt^4$ jets separately, they can be compared directly to signal jets of similar jet substructure (as discussed in Section~\ref{subsec:Calibration_of_signal}) and systematic uncertainties can be derived as discussed in Section~\ref{sec:uncertainties_for_Signal}.
For the calibration in data, these categories are difficult to distinguish experimentally and their tagger response is similar.
Thus, $\PQt^3$ and $\PQt^4$ jets are treated together and designated $\PQt^{3,4}$ in the following.
In simulation, jets are placed into these categories, as well as signal categories, by matching the reconstructed jets to the generator-level partons in $\Delta R$. 
The matching criteria are summarized in Table~\ref{tab:Matching}.
The proportion of jets not matched to any of these categories, is less than 6 (5)\% of the SR (preselection) events, and they have a negligible impact on the analysis.

\begin{table*}[htb]
\centering
\topcaption{
Matching criteria used to place a jet in one of the SM jet categories (left four columns) or merged radion jet categories (right two columns).
Each column lists the $\Delta R$ conditions demanded between the reconstructed jet (j) and the generator-level parton in order to match a jet with a particular jet substructure.
Lower indexes enumerate partons and indicate the particle from whose decay they originate (\eg, $\PQt \to \PQb_\PQt\PQq_{1\PW}\PQq_{2\PW}$).
Schematic diagrams for each jet type are shown below each column.}
\cmsTable{
\begin{scotch}{cccccc}
\qg                   & \PW                            & $\PQt^2$                      & $\PQt^{3,4}$                  &  $\PR^{3,4\PQq}$            & \Rlqq      \\\hline
$(\qg,\text{j})< 0.6$ & $(\PW,\text{j})< 0.6$          & $(\PQt,\text{j})< 0.6$        & $(\PQt,\text{j})< 0.6$        &  $(R,\text{j})     <0.6$    & $(R,\text{j})     <0.6$  \\ [\cmsTabSkip]
\NA                   & $(\PQq_{1\PW},\text{j})< 0.8$  & $(\PQb_\PQt,\text{j})< 0.8$   & $(\PQb_\PQt,\text{j})< 0.8$   &  $(\PQq_{1},\text{j}) <0.8$ & $(\PQq_{1},\text{j}) <0.8$ \\
\NA                   & $(\PQq_{2\PW},\text{j})< 0.8$  & $(\PQq_{1\PW},\text{j})< 0.8$ & $(\PQq_{1\PW},\text{j})< 0.8$ &  $(\PQq_{2},\text{j}) <0.8$ & $(\PQq_{2},\text{j}) <0.8$ \\
\NA                   & $(\PQb_\PQt,\text{j})> 0.8$    & $(\PQq_{2\PW},\text{j})> 0.8$ & $(\PQq_{2\PW},\text{j})< 0.8$ &  $(\PQq_{3},\text{j}) <0.8$ & $(\Pell,\text{j})  <0.8$  \\
\multirow{2}{*}{\NA}  & \multirow{2}{*}{\NA}           & \multirow{2}{*}{\NA}          & For $\PQt^4$ ($\PQt^3$):    &  For \Rfourq (\Rthreeq)  :  & \multirow{2}{*}{\NA} \\
                      &                                &                               & $(\qg,\text{j}) < (>)\,0.8$  &  $(\PQq_{4},\text{j})<(>)\,0.8$ &     \\
\end{scotch}
}
\includegraphics[width=\textwidth]{Figure_004-a.pdf}
\label{tab:Matching}
\end{table*}

The calibration of the \PW, $\PQt^2$, and $\PQt^{3,4}$ jets requires samples enriched in those jets.
Therefore, dedicated calibration samples are defined,
and the calibration for these jets is summarized in Section~\ref{subsec:Calibration_of_W_t}.
The \qg jets are calibrated using preselection jets, and this procedure is described in Section~\ref{subsec:Calibration_of_q/g}.
The calibration of signal jets is presented in Section~\ref{subsec:Calibration_of_signal}.

\subsection{Calibration of \texorpdfstring{\PW}{W} boson and top quark jets with a matrix method} \label{subsec:Calibration_of_W_t}

For the calibration of the taggers, a control sample similar to the preselected one, but enriched in \PW boson and top quark jets, is used. We refer to this sample as the ``sideband''.
The sideband is defined by requiring one isolated lepton (\PGm or \Pe), $\ptmiss > 40 (80)\GeV$ for \PGm (\Pe), and one or two massive jets.
The neutrino $p_z$ is reconstructed under the assumptions that the invariant mass of the $\Pell\PGn$ system is equal to the \PW boson mass $m_{\PW}=80\GeV$ (as described in Ref.~\cite{CMS:2021klu}) and the transverse momentum of the $\Pell\PGn$ system is required to satisfy $\pt^{\Pell\PGn}>200\GeV$.
This means that for the sideband one of the massive jets used for the preselection is effectively replaced by a leptonically decaying \PW boson candidate.

The highest mass jets in these sideband events with $\MJ>60\GeV$ are used for the calibration.
These jets are categorized by the matching to the \PW, $\PQt^2$, $\PQt^{3,4}$, and \qg categories described previously.
We split the events into two \MJ bins, one with $60 < \MJ < 120\GeV$ (low mass) and the other one with $\MJ > 120\GeV$ (high mass).
In addition, we split the sideband events further into two bins, based on the jet \ptj. For low-mass jets, the bins used are 200--400 and $>$400\GeV, while for high-mass jets the bins used are 200--500 and $>$500\GeV.
The resulting four samples are designated LL, LH, HL, and HH, where the first letter indicates low or high \MJ and the second letter low or high \ptj.
The two SM processes (\Wjets and top quark production) are normalized in each of these four categories by scaling them to match the data separately for events with zero or one \PQb-tagged jet(s).
This corrects the simulation for an $\mathcal{O}$(10\%) mismodeling of the cross section, and the residual data-to-simulation differences in the \dPWorWH distribution can be attributed to the mismodeling of the discriminant.

The LL and LH samples contain primarily events of the \PW, $\PQt^2$, and \qg jet categories; the HL and HH samples primarily events of the $\PQt^2$, $\PQt^{3,4}$, and \qg jet categories.
Any other jet contribution or unmatched jets (collectively $<$5\%) can be ignored.
For each of these jet types, we apply a set of kinematic conditions to split them further into three subsamples so that each sample is highly pure in a single jet type.
The splitting conditions include kinematic variables such as $N$-subjettiness~\cite{Thaler_2012}, \NJ, \NB, and \MJ, as well as \textsc{DeepAK8} discriminants other than the calibrated one.

\begin{figure*}[ht!]\centering
\includegraphics[width=0.49\linewidth]{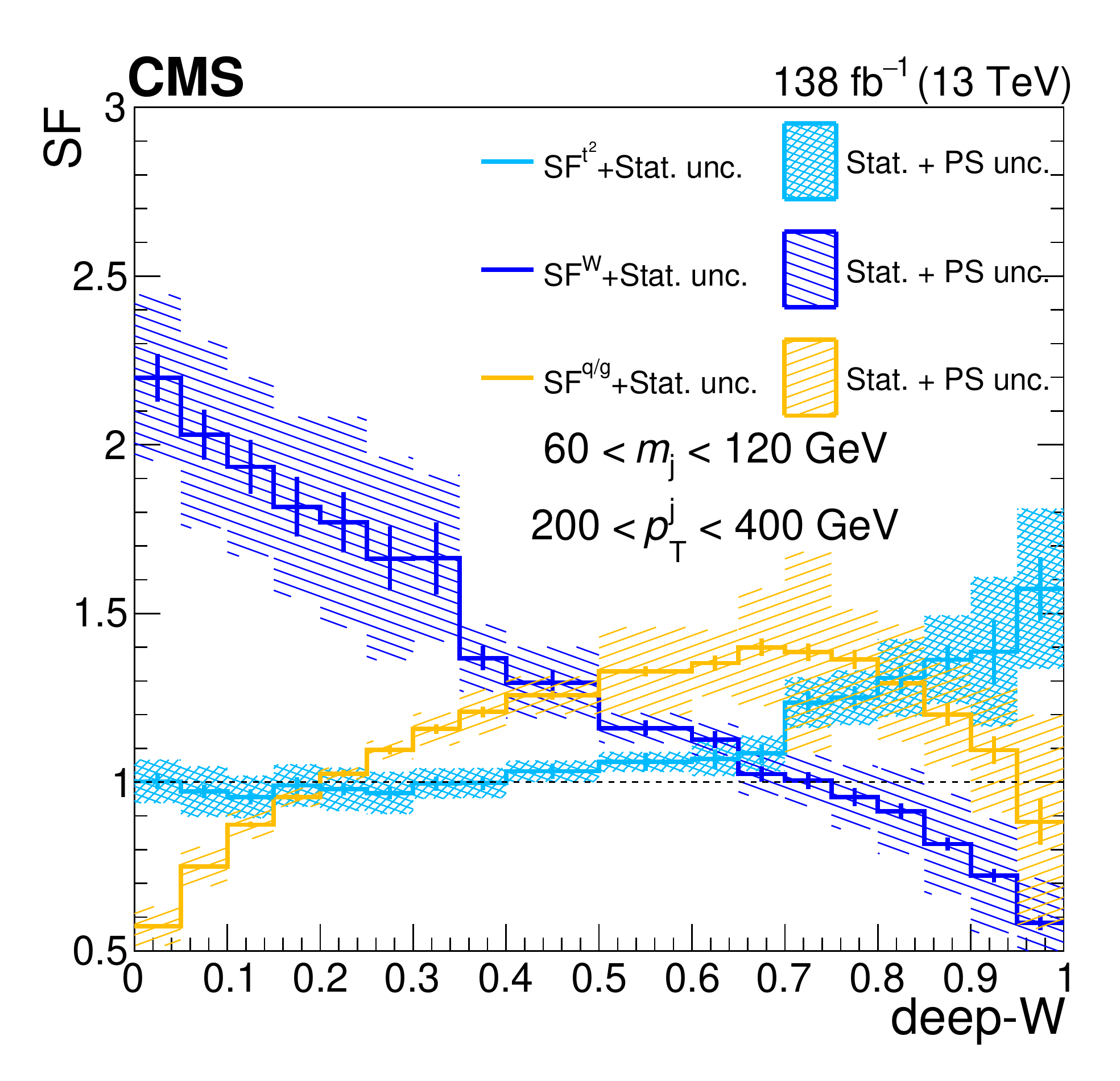}
\includegraphics[width=0.49\linewidth]{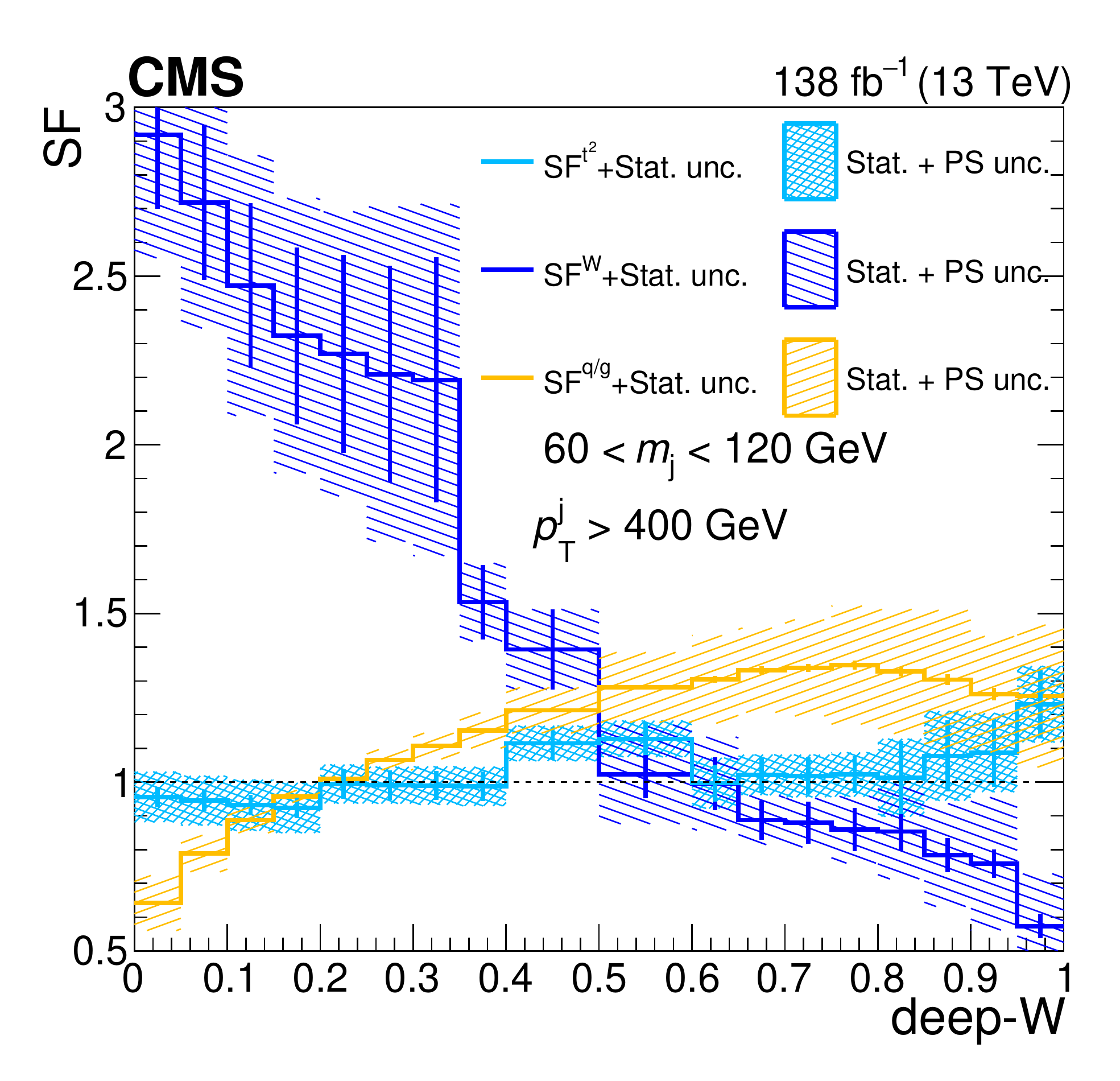}
\includegraphics[width=0.49\linewidth]{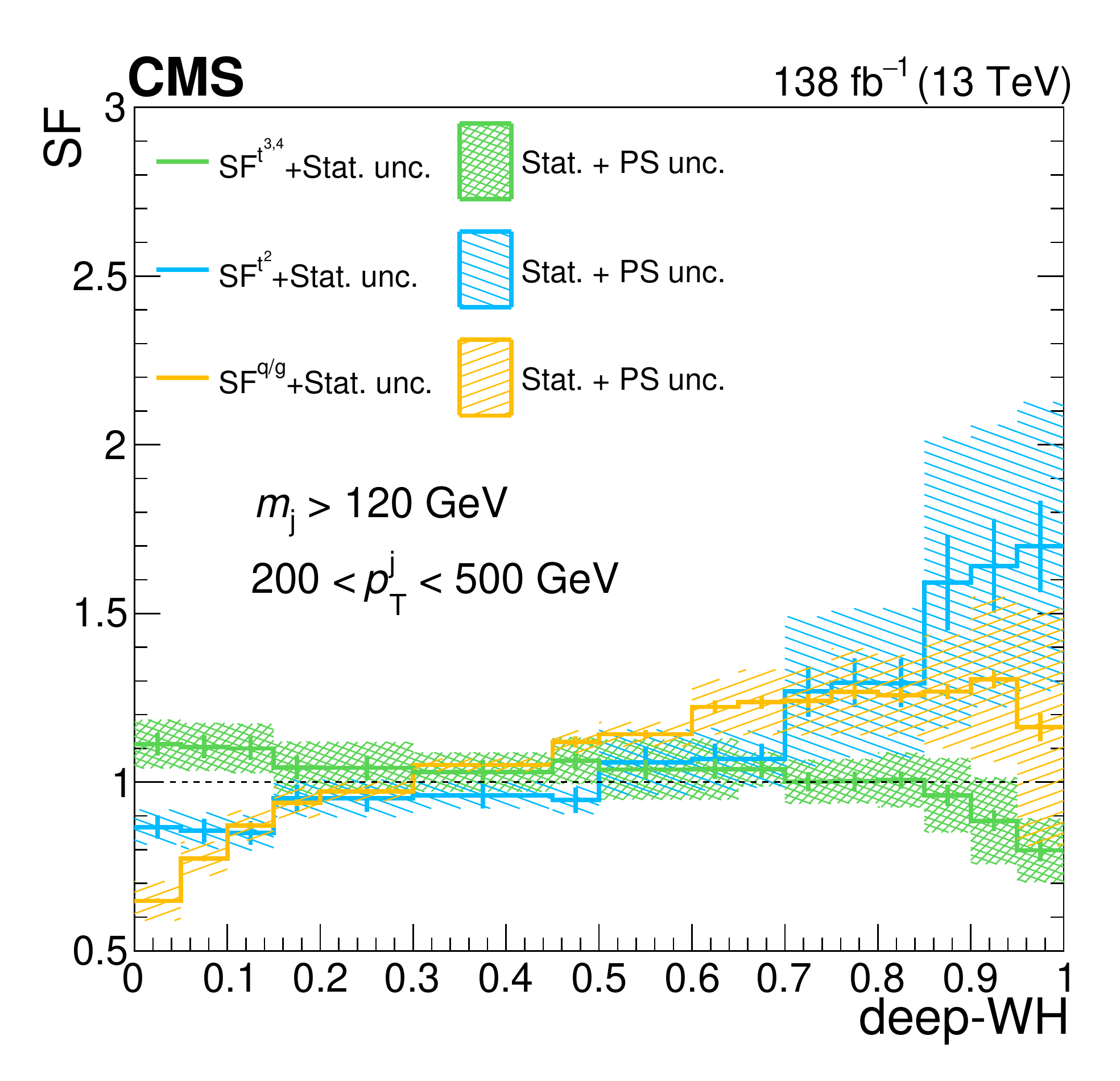}
\includegraphics[width=0.49\linewidth]{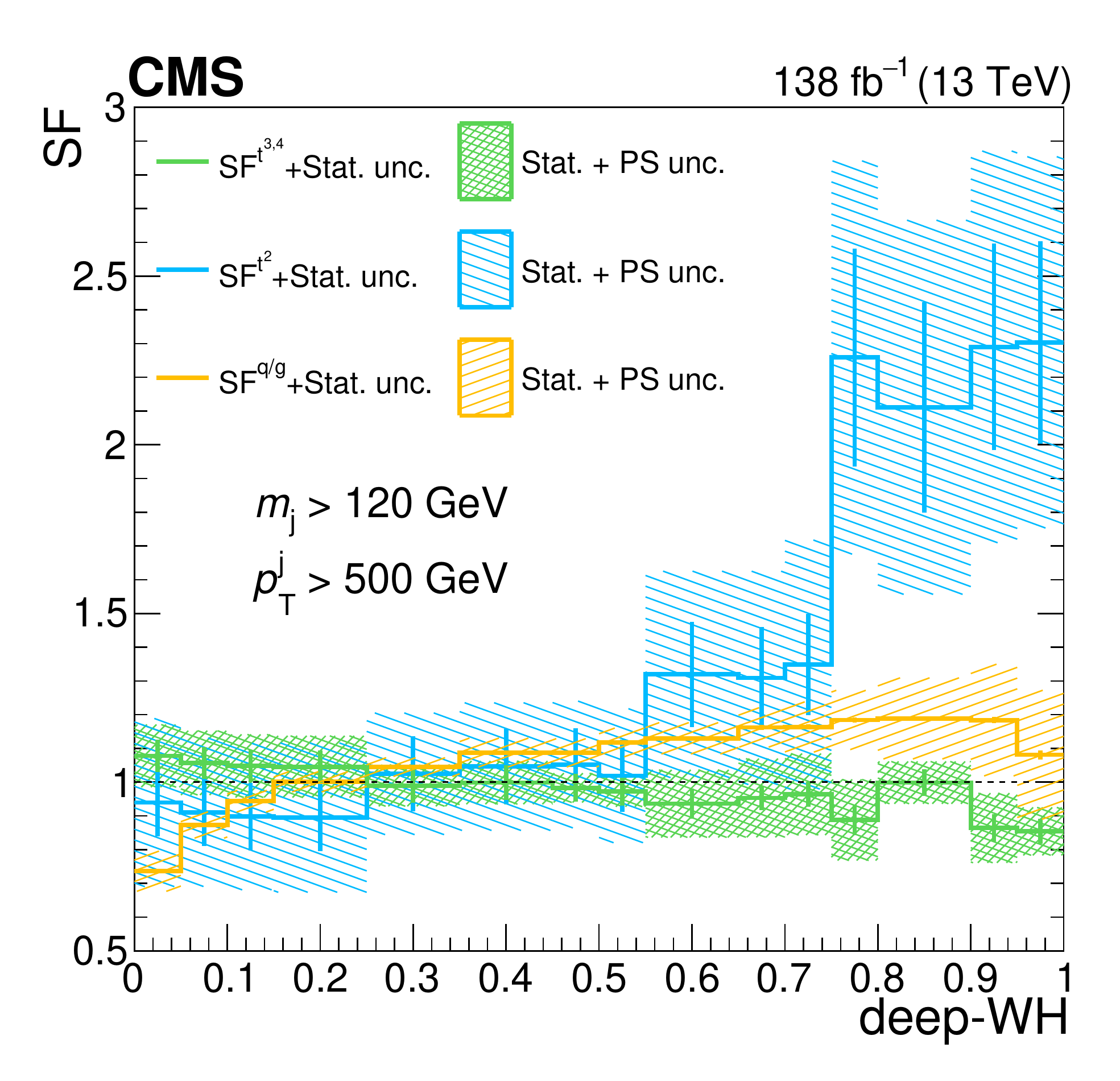}
\caption{
Measured scale factors (SFs) for the \dPW and $\dPW\PH$ discriminants.
Upper row: SFs for \PW (dark blue), $\PQt^2$ (light blue), and \qg (yellow) matched jets in the low-\MJ bins, LL (left) and LH (right), as functions of the \dPW discriminant value.
Lower row: SFs for $\PQt^2$ (light blue), $\PQt^{3,4}$ (green), and \qg (yellow) matched jets in the high-\MJ bins, HL (left) and HH (right), as functions of the $\dPW\PH$ discriminant value.
For each discriminant value bin, the sum of the SF-corrected jet yields is required to be equal to the observed data.
The statistical and parton shower (PS) uncertainties are shown by the shaded bands.
}
\label{fig:SFs}
\end{figure*}

The \dPWorWH distributions are formed for each of the three pure subsamples for LL.
One equation is written for each pure subsample by equating the data yields $D_{i,k}$ for a jet type $i$ in a \dPWorWH bin $k$ with the simulated jet yields for \PW, $\PQt^2$, and \qg (which we write as \PW, \PQt, and \Pg here), scaled by the scale factors $\text{SF}_{k}^\PW$, $\text{SF}_{k}^\PQt$, and $\text{SF}_{k}^\Pg$:
$D_{i,k} = \text{SF}_{k}^\PW W_{i,k} + \text{SF}_{k}^\PQt t_{i,k} + \text{SF}_{k}^\Pg g_{i,k} + d_{i,k}$.
The $d_{i,k}$ term accounts for the other types of jet yields; their contribution is small (amounting to $<$5\% for most of the bins), and these jet types are treated as not contributing to the mismodeling.
A similar equation can be written for each of the three ($i = 1, 2, 3$) subsamples \PW, \PQt, and \Pg to form a system of three equations:
\begin{linenomath}
\begin{equation}
\begin{pmatrix}
D_{1,k} - d_{1,k} \\
D_{2,k} - d_{2,k} \\
D_{3,k} - d_{3,k}
\end{pmatrix}=
\begin{pmatrix}
W_{1,k} & t_{1,k} & g_{1,k}\\
W_{2,k} & t_{2,k} & g_{2,k}\\
W_{3,k} & t_{3,k} & g_{3,k}
\end{pmatrix}
\begin{pmatrix}
\text{SF}^\PW_k \\
\text{SF}^\PQt_k \\
\text{SF}^\Pg_k
\end{pmatrix},
\end{equation}
\end{linenomath}
in which the jet yields and the data are known, while the three SFs ($\text{SF}_{k}^\PW$, $\text{SF}_{k}^\PQt$, $\text{SF}_{k}^\Pg$) are unknown.
We solve this $3{\times}3$ system per \dPWorWH bin $k$ to derive the SFs for each type of jet.
The scale factors obtained with this matrix method $\text{SF}_{k}^\PW, \text{SF}_k^{\PQt^2}$ and $\text{SF}_k^{\PQt^{3,4}}$ are shown in Fig.~\ref{fig:SFs} for LL, LH, HL, and HH.
As the three subsamples are highly enriched in exactly one jet type, the matrix is nearly diagonal, and the derivation of the SFs is dominated by the data vs. simulation modeling in the corresponding pure subsamples.
For example, the data/simulated yields in the \PW-pure subsample dominate the determination of $\text{SF}_{k}^\PW$.
The method yields reliable SFs in the regime where subsamples are highly pure.
Both \dPW and $\dPW\PH$ are calibrated with this procedure for each of the LL, LH, HL, and HH bins separately.
While the SFs are quite large and vary from about 0.5 to 3, the integral over the tagger score yields an effective SF close to 1;
for example, \PW-boson jets with $\dPW>0.6\,(0.8)$ have effective SFs of 0.89 (0.78) and 0.80 (0.74) for the LL and LH samples, respectively.

\begin{figure*}[htbp]
\centering
\includegraphics[width=0.48\textwidth]{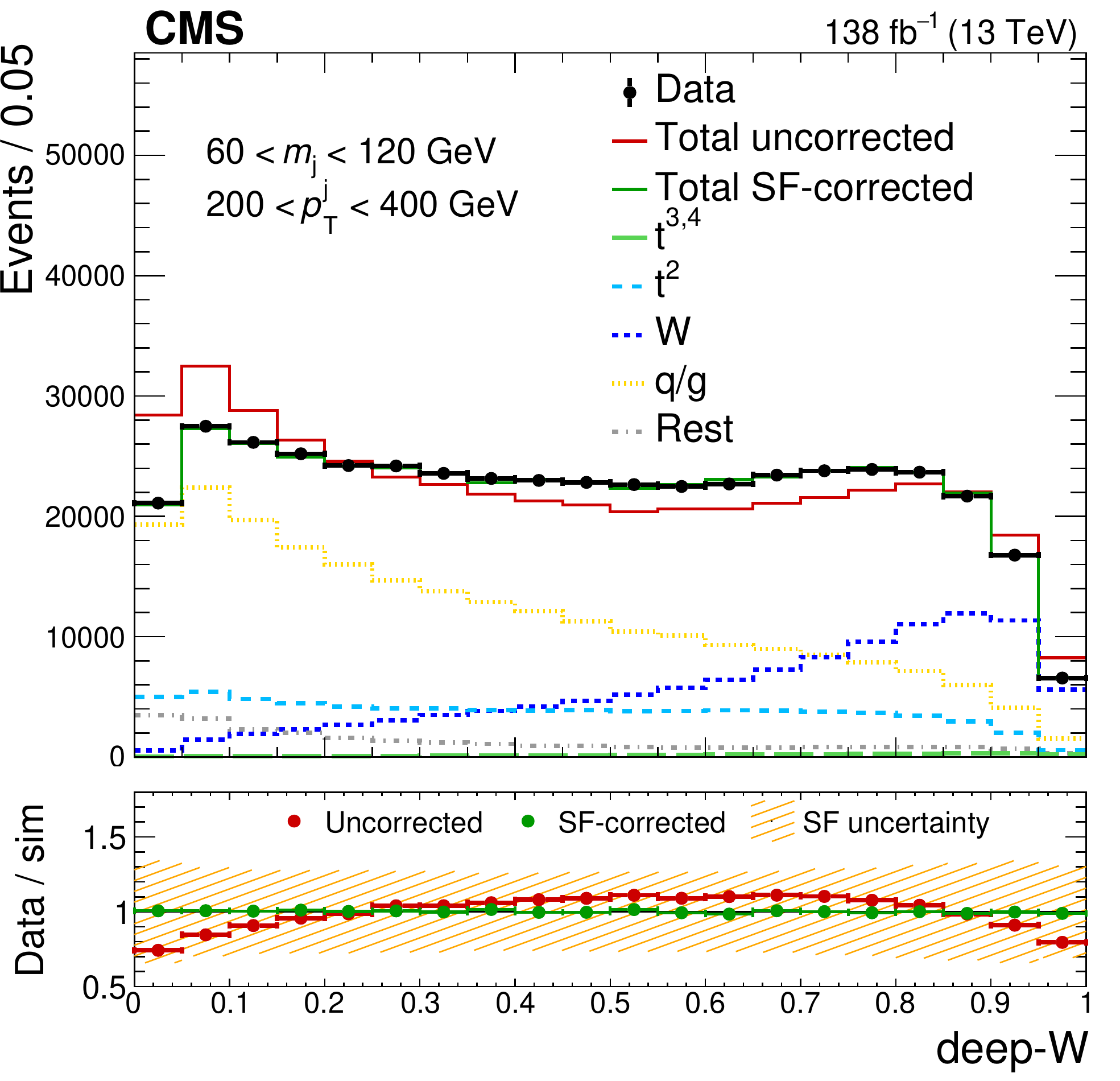}
\includegraphics[width=0.48\textwidth]{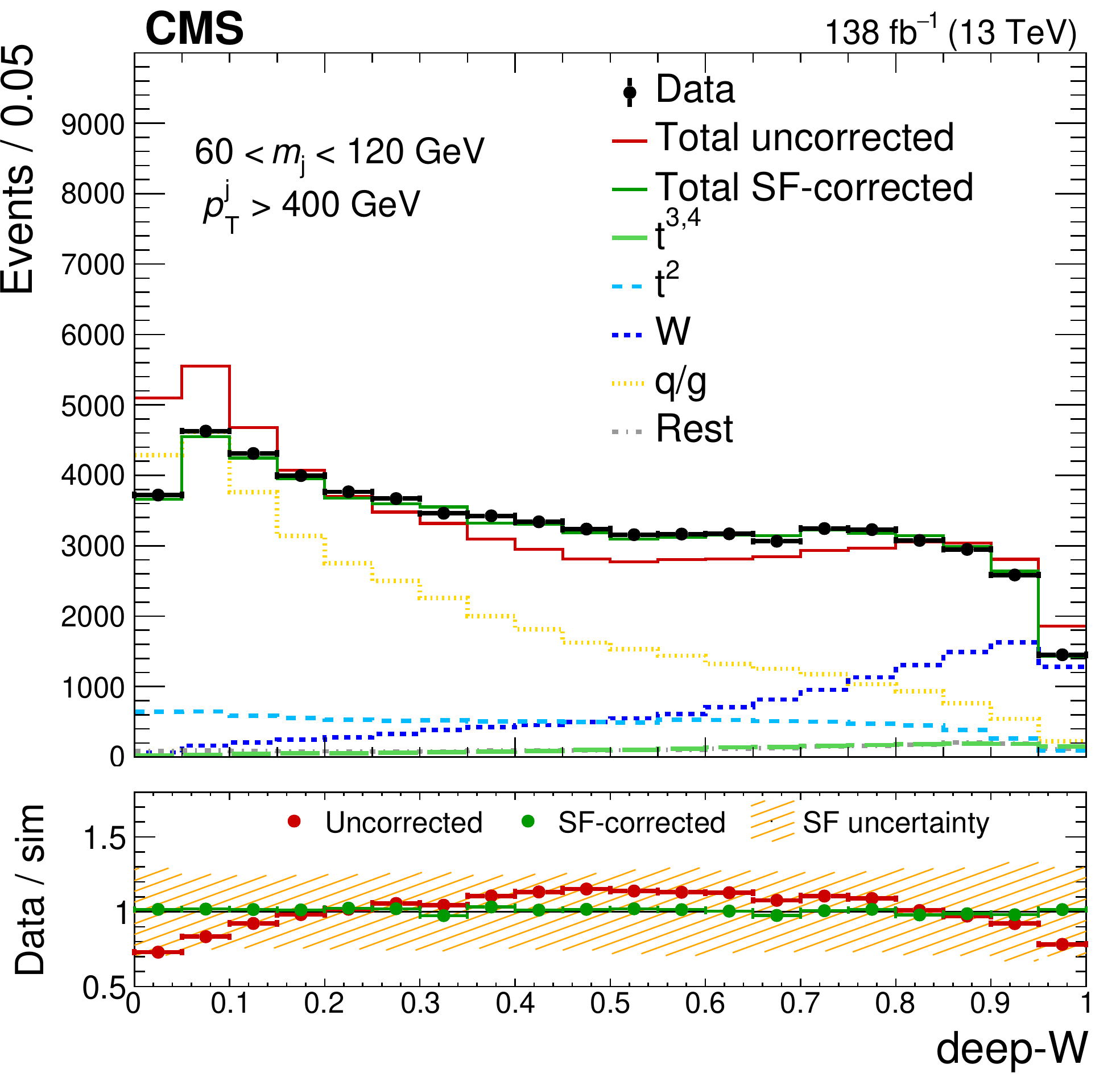}
\includegraphics[width=0.48\textwidth]{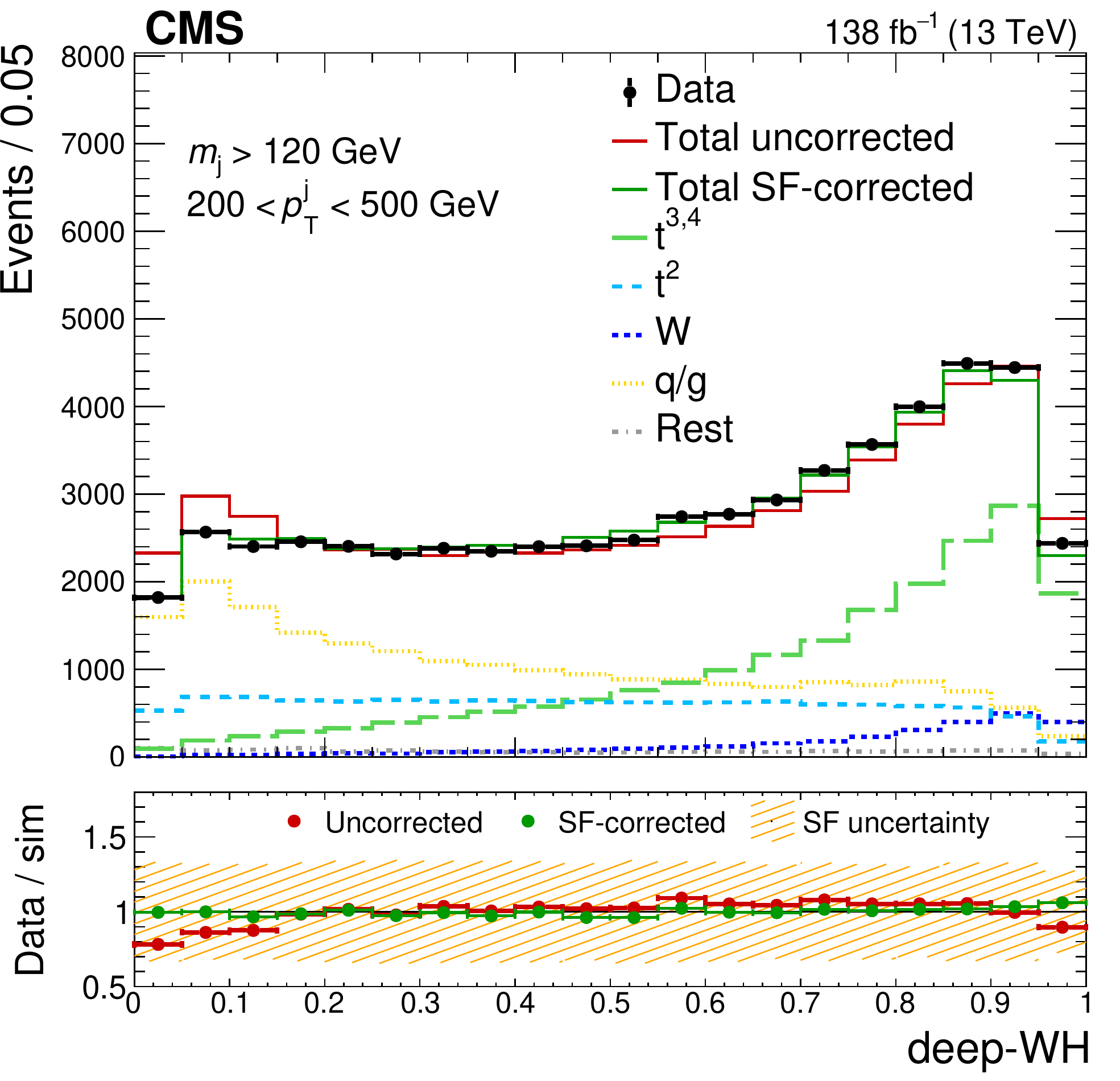}
\includegraphics[width=0.48\textwidth]{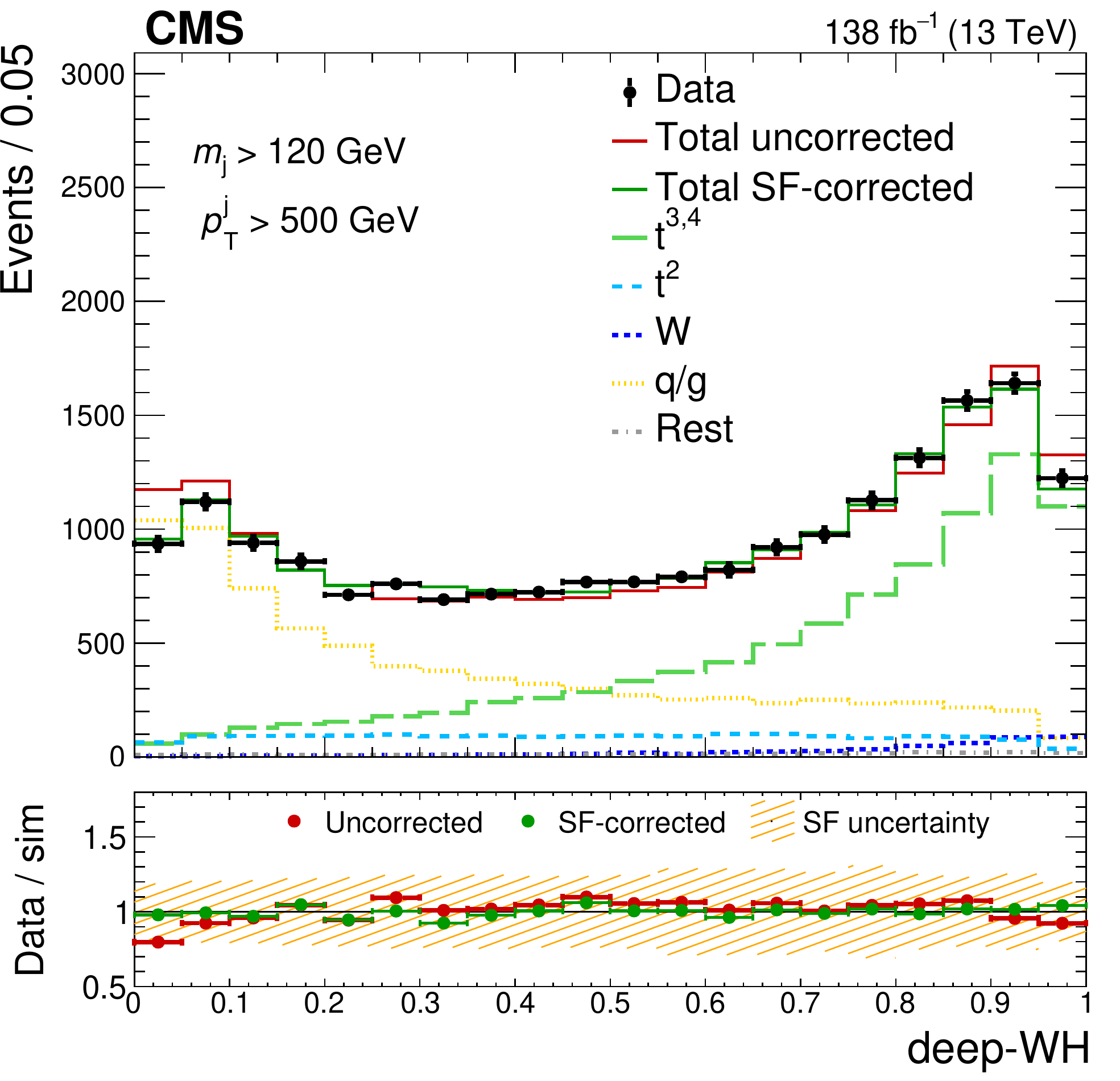}
\caption{
\textsc{DeepAK8} discriminants of the jet with highest mass in the single-lepton sideband.
The \dPW spectra in the LL (upper left) and LH (upper right) samples are presented together with
the $\dPW\PH$ spectra in the HL (lower left) and HH (lower right) samples.
The \PW boson jets are shown in dark blue, $\PQt^2$ in light blue, $\PQt^{3,4}$ in green, \qg in yellow,
and the ``Rest'' jet types (jets not matching any of the categories) in gray.
Before corrections (red), discrepancies between the prediction and the data can be observed, in particular at low and high discriminant values.
The corrected distributions after application of the scale factors (SFs) are shown in dark green.
The lower panels show the data-to-simulation ratios before and after corrections.
The SF uncertainties are indicated by the shaded bands.}
\label{fig:decomp}
\end{figure*}

All simulated events, based on the types of the selected jets they contain and their \ptj and \MJ, are corrected by the SFs for the respective \dPWorWH bins.
The discriminant distributions before and after corrections are shown in Fig.~\ref{fig:decomp}.
Various validation tests show good agreement between data and simulation.
As the extracted SFs are found to depend on the choice of splitting conditions defining the pure subsamples,
systematic uncertainties resulting from the selection criteria are assigned, as described in Section~\ref{sec:uncertainties}.

\subsection{Calibration of quark and gluon jets} \label{subsec:Calibration_of_q/g}

The quark and gluon jets are treated collectively as a single type of jet, \qg.
Their calibration is performed using the preselected sample where SR events and events with \PQb-tagged AK4 jets are vetoed.
This sample consists of more than 13 million events, of which more than 97\% are QCD multijet events.
Similarly to the single-lepton sideband sample, we consider only the highest mass jet with $\MJ > 60\GeV$ in each event, and define the same four LL, LH, HL, and HH bins in \MJ and \ptj.
The QCD events in each bin are normalized to the data.
The contribution from \PW, $\PQt^2$, and $\PQt^{3,4}$ jets, amounting to less than 2\%, is estimated using simulation and subtracted from the data. The result is divided by the \qg yields to define $\text{SF}^{\qg}_k$ in each \dPWorWH discriminant value bin $k$.
The resulting values of $\text{SF}^{\qg}$ are presented together with $\text{SF}^\PW$, $\text{SF}^{\PQt^2}$, and $\text{SF}^{\PQt^{3,4}}$ in Fig.~\ref{fig:SFs}.
The relative fraction of quarks and gluons is the same for the preselection region where the SFs are defined, and the SRs and control regions (CRs).
The only difference between the jets are therefore their \pt spectra.

\begin{figure*}[htbp]\centering
\includegraphics[width=0.42\linewidth]{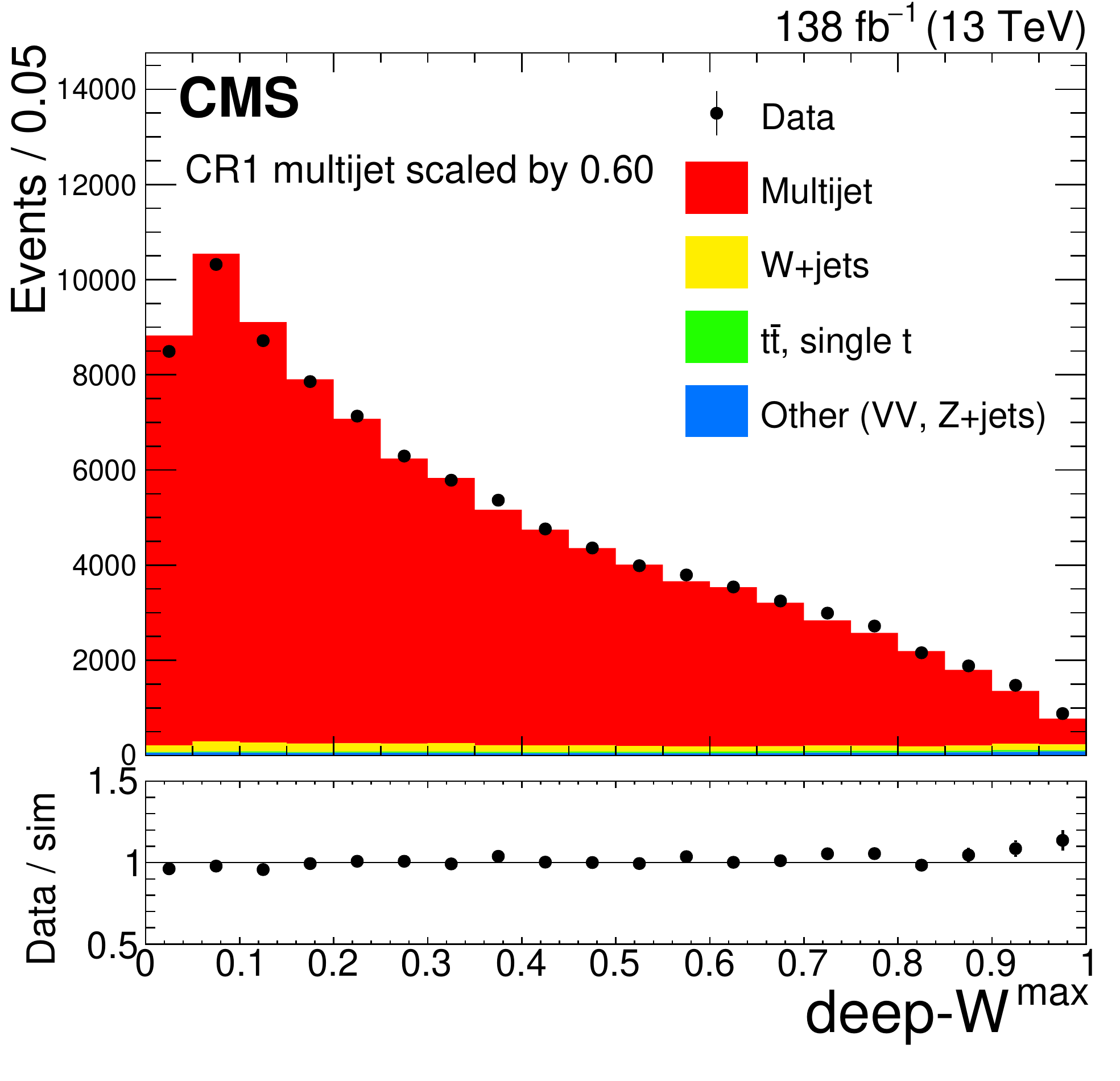}
\includegraphics[width=0.42\linewidth]{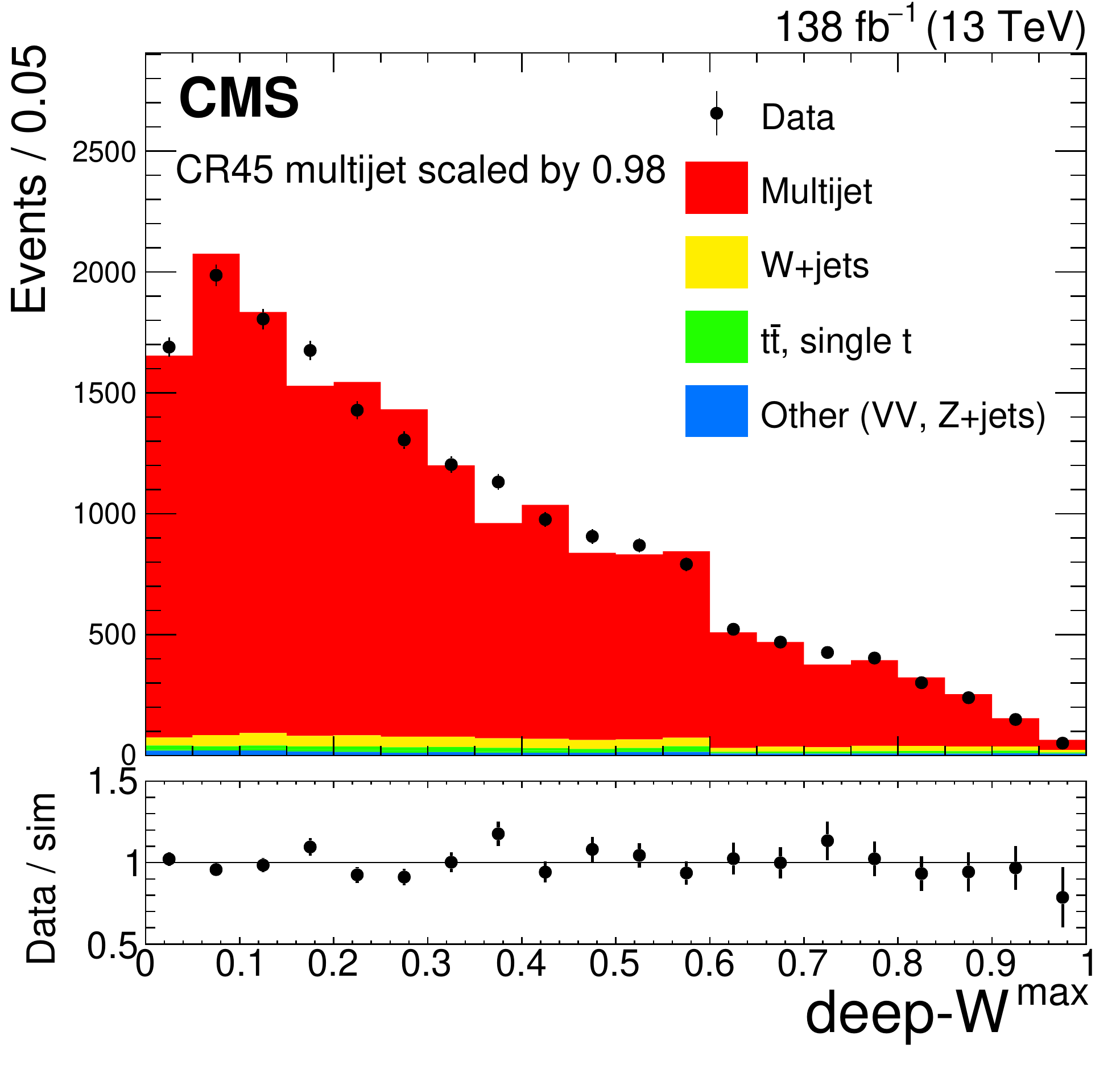}
\includegraphics[width=0.42\linewidth]{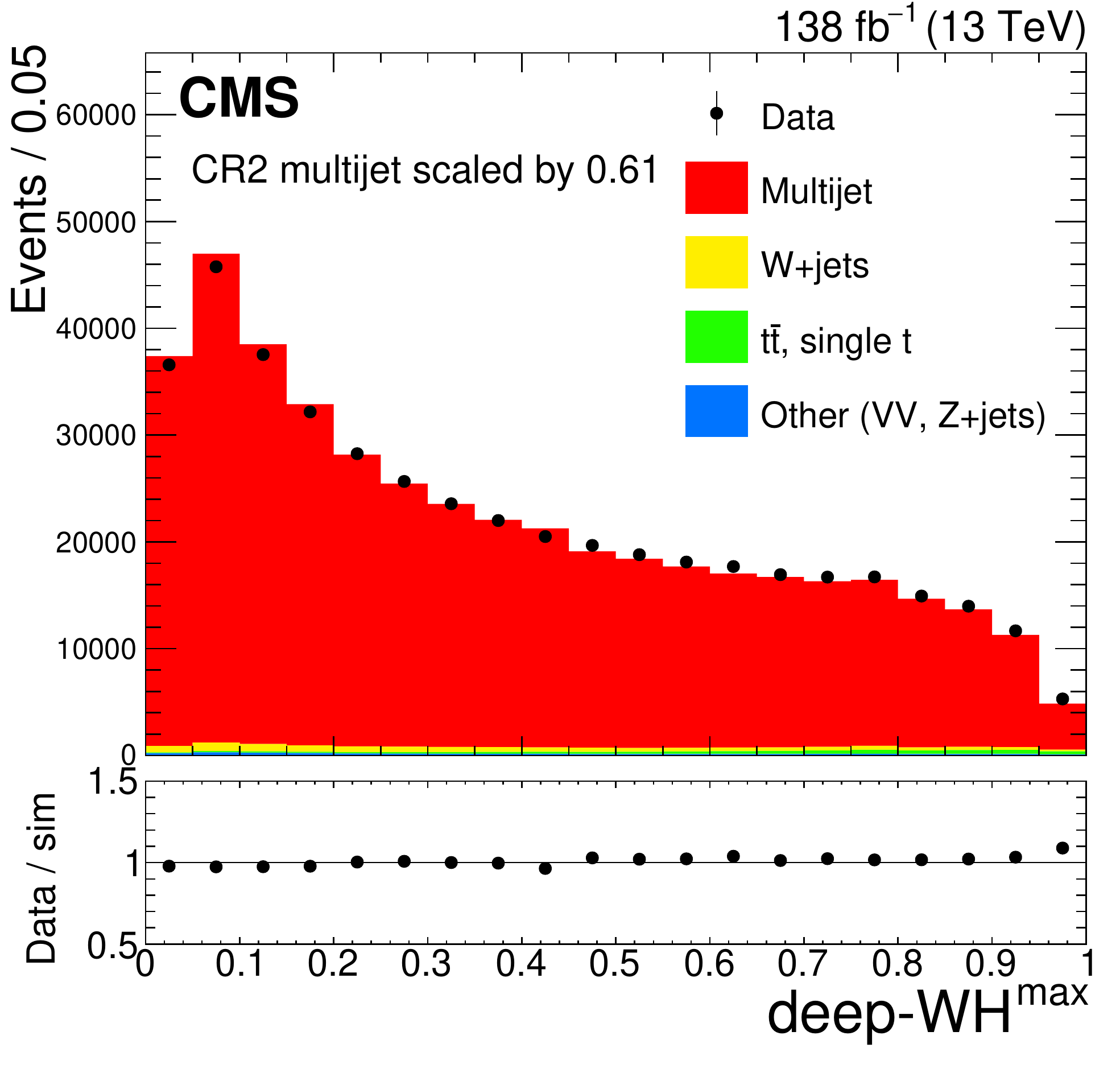}
\includegraphics[width=0.42\linewidth]{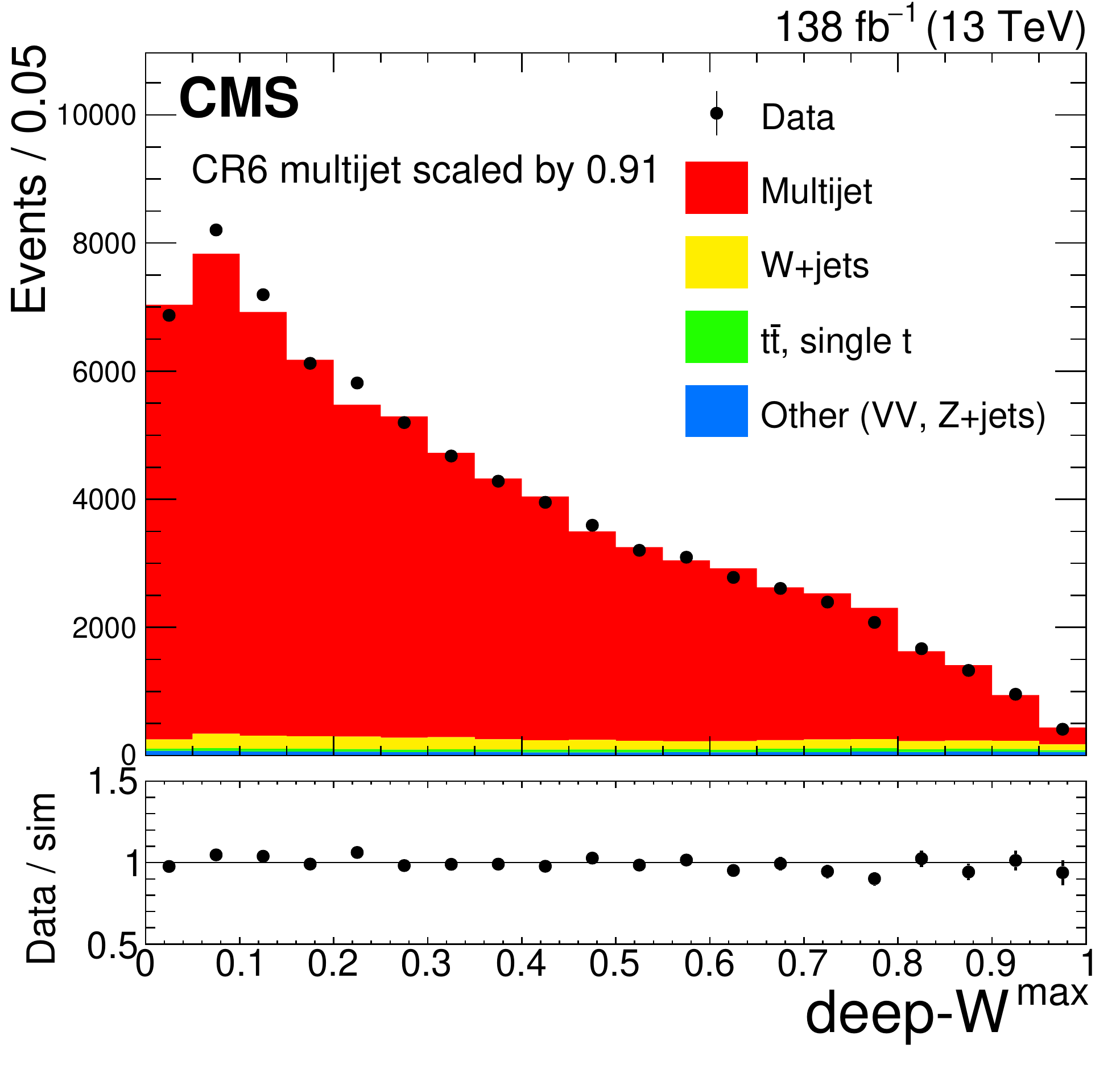}
\raggedright
\includegraphics[width=0.42\linewidth]{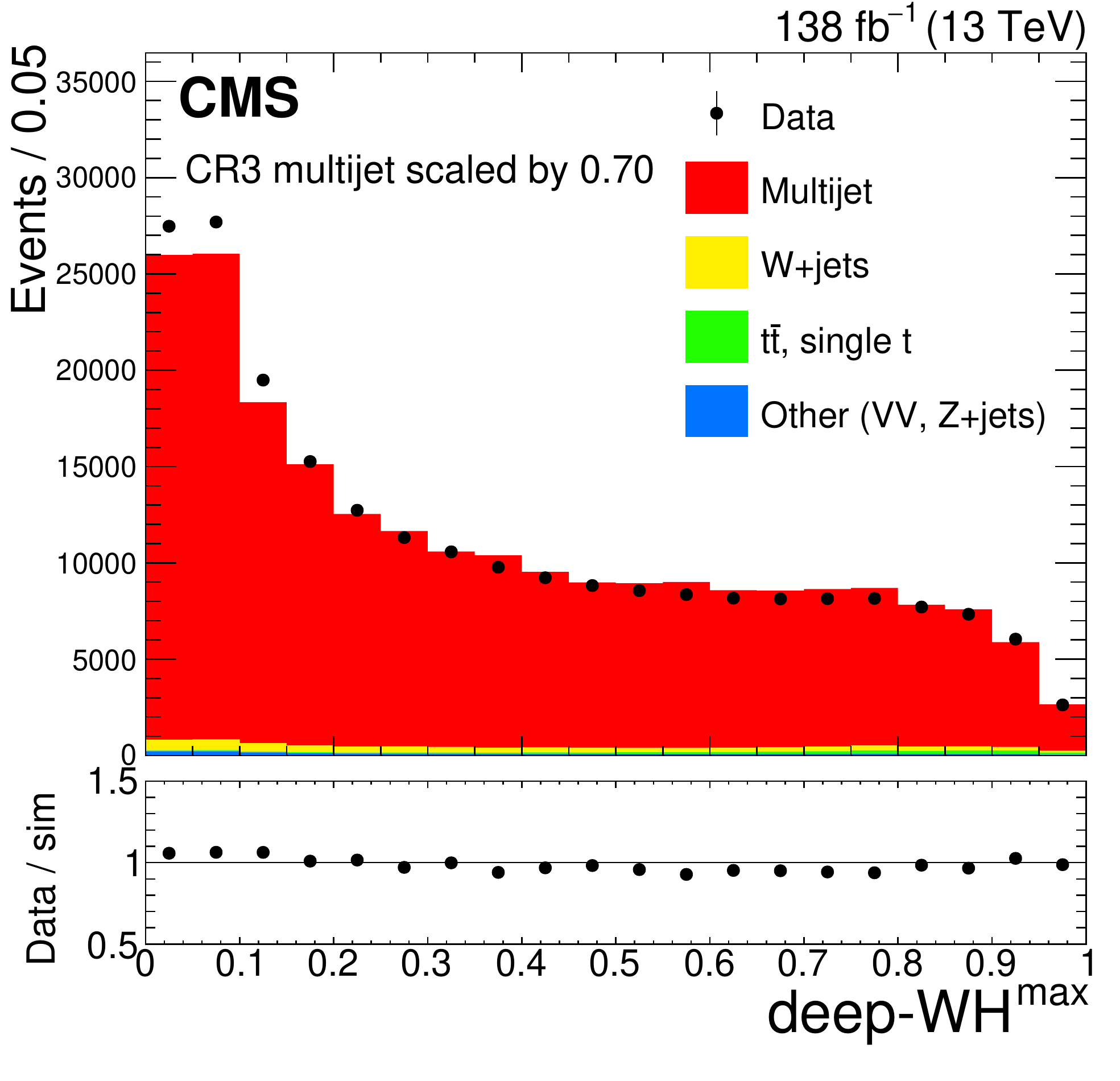}
\caption{Comparison between data (black markers) and simulated background events (histograms) of the \dPWorWH distributions for the highest mass jet after SF application.
The control regions CR1, CR2, CR3 are shown in the left column, upper to lower rows, while CR45 and CR6 are presented in the right column, upper and middle rows, respectively.
The lower panels show the data-to-simulation ratio.}
\label{fig:CRs_deepScores_SF_Corrected}
\end{figure*}

Validation tests have shown a good post-correction performance, where the ratio of data to simulation is consistent with unity over the entire \dPWorWH range.
To perform these tests, we define CRs by using the SR1--6 selections with at least one of the \dPWorWH conditions inverted.
For SR4 and SR5, this inversion leads to the same sample with zero or one \PW-tagged jet.
The five resulting CRs, associated with the SRs, are named CR1, CR2, CR3, CR45, and CR6.
Figure~\ref{fig:CRs_deepScores_SF_Corrected} shows the \dPWorWH distributions for the highest mass jet in each CR after the SFs are applied.
A similar, almost flat performance is exhibited by the middle and minimum mass jets.
Validation tests in samples using other CR definitions lead to similar post-correction performance.

\subsection{Calibration of signal jets with SM proxy jets} \label{subsec:Calibration_of_signal}

The \dPWorWH discriminant distributions for simulated signal events are also corrected using SFs.
For resolved-radion signal events ($\NJ = 3$, SR4--6), the \PW-boson-matched jets are scaled by $\text{SF}^\PW$
according to the \ptj, \MJ, and \dPW values for each jet.

\begin{figure}[htp!]\centering
\includegraphics[width=\triFigWidth]{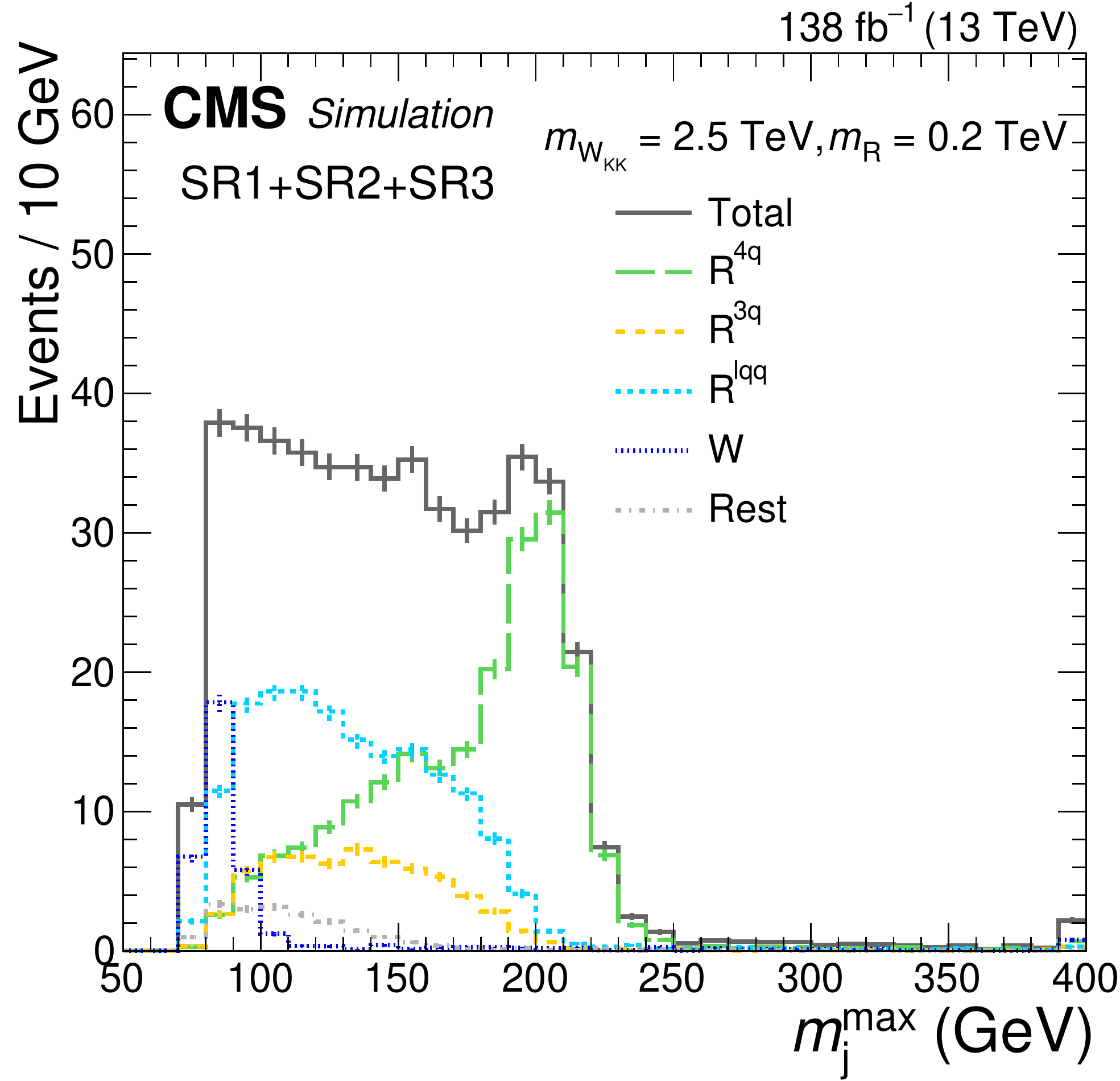}
\includegraphics[width=\triFigWidth]{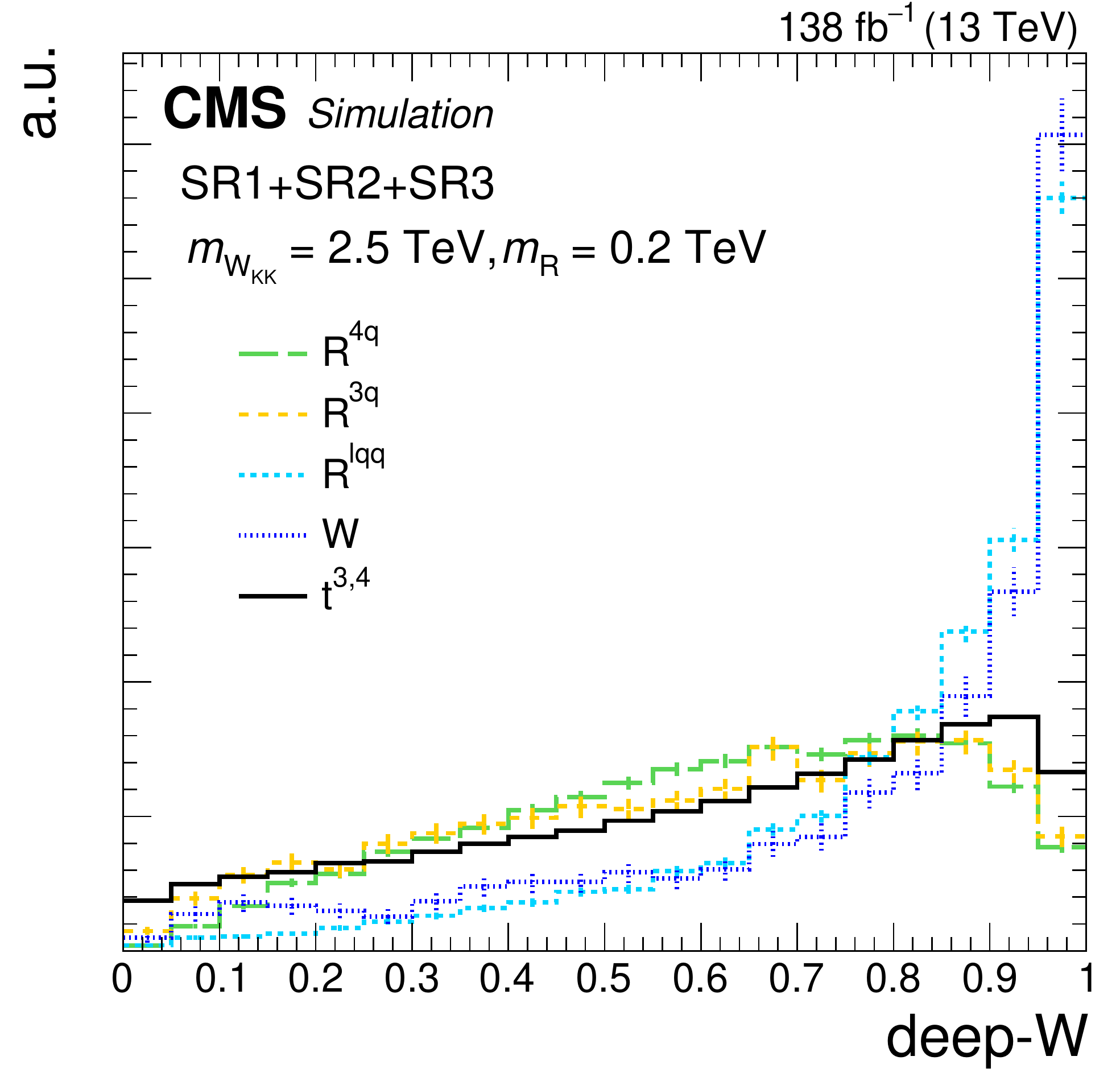}
\includegraphics[width=\triFigWidth]{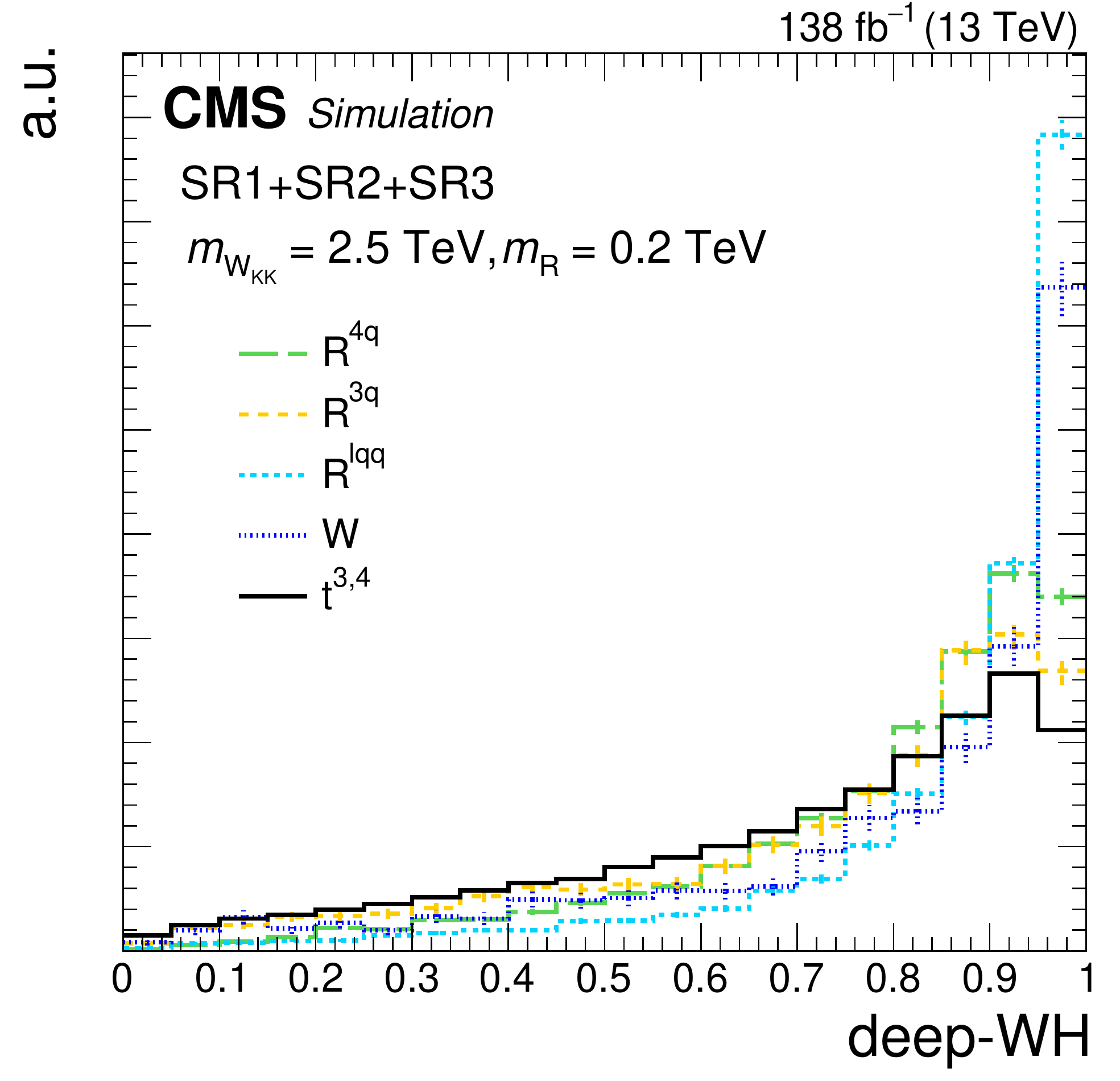}
\caption{
Shape comparison for different jet types in simulation.
\shapeFigLeft: the \MJmax distributions for SR1--3 events without \dPWorWH constraints.
\shapeFigMiddle and \shapeFigRight: the \dPW and $\dPW\PH$ distributions normalized to unity for the shown components, respectively.
The $\PQt^{3,4}$ jets from the preselected sample, normalized to unity, are superimposed to compare shapes with the \Rthreeq and \Rfourq distributions.}
\label{fig:SignalDecomposition}
\end{figure}

Merged radion signal events ($\NJ = 2$, SR1--3) contain jets of the form \PW, \Rlqq, \Rthreeq, and \Rfourq.
Figure~\ref{fig:SignalDecomposition} (\ifthenelse{\boolean{cms@external}}{upper}{left}) shows the relative contributions of each of these categories to the total as a function of \MJmax.
There are very few SM jets with the same substructure and flavor compositions as \Rlqq, \Rthreeq, and \Rfourq jets that can be directly used for calibration (considering that boosted Higgs bosons are not abundant).
Instead, we calibrate these using SM jets that have similar prong substructure and \dPWorWH response, which we call ``proxy jets''.

The \PW boson jets exhibit highly similar \dPW and $\dPW\PH$ distributions to \Rlqq jets.
Thus, we use \PW boson jets as proxy jets for the \Rlqq calibration.
The similarity of the two spectra can be seen in Fig.~\ref{fig:SignalDecomposition} for both the \dPW (used in SR1) and $\dPW\PH$ (used in SR2--3) discriminants.
This similarity results from the discriminant design, as the raw scores in both the numerator and denominator have not been derived for events with leptons inside jets, and so the \dPW and $\PW\PH$ discriminants are largely blind to the presence of a lepton.

The closest abundant SM jets with substructure similar to \Rthreeq and \Rfourq are fully merged top quark jets $\PQt^{3,4}$.
As Fig.~\ref{fig:SignalDecomposition} (\ifthenelse{\boolean{cms@external}}{lower}{right}) shows, the $\dPW\PH$ distributions of those three jet types are similar and thus the $\PQt^{3,4}$ jets are used as proxy jets to calibrate signal \Rthreeq and \Rfourq jets.
Accordingly, the corresponding $\text{SF}^{\PQt^{3,4}}$ values derived in Section~\ref{subsec:Calibration_of_W_t} are used to calibrate \Rthreeq and \Rfourq jets.
We find that the individual $\PQt^3$ and $\PQt^4$ components have an even better shape agreement with their corresponding signal jets \Rthreeq and \Rfourq, respectively.
This consistency suggests that despite their differences (in quark flavor, kinematics, and color recombination), the $\PQt^{3,4}$ and $\PR^{3\PQq,4\PQq}$ jets have a largely similar response to the $\dPW\PH$ discriminant.
Systematic uncertainties are assigned to account for differences in this response and also to account for residual shape differences as discussed in Section~\ref{sec:uncertainties}.

\section{Background estimation} \label{sec:bkg_estimation}

\begin{figure*}[htbp]\centering
\includegraphics[width=0.42\textwidth]{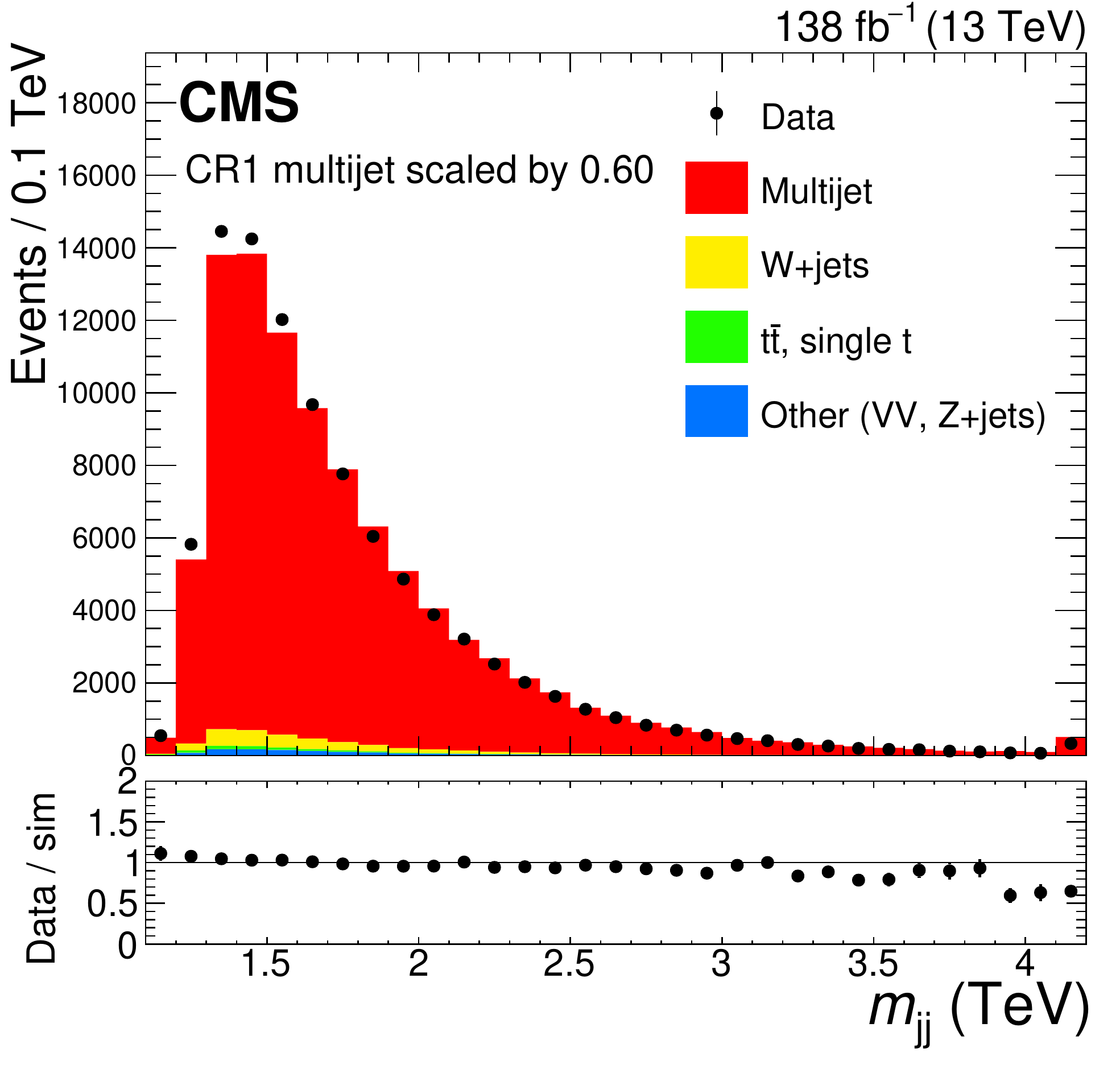}
\includegraphics[width=0.42\textwidth]{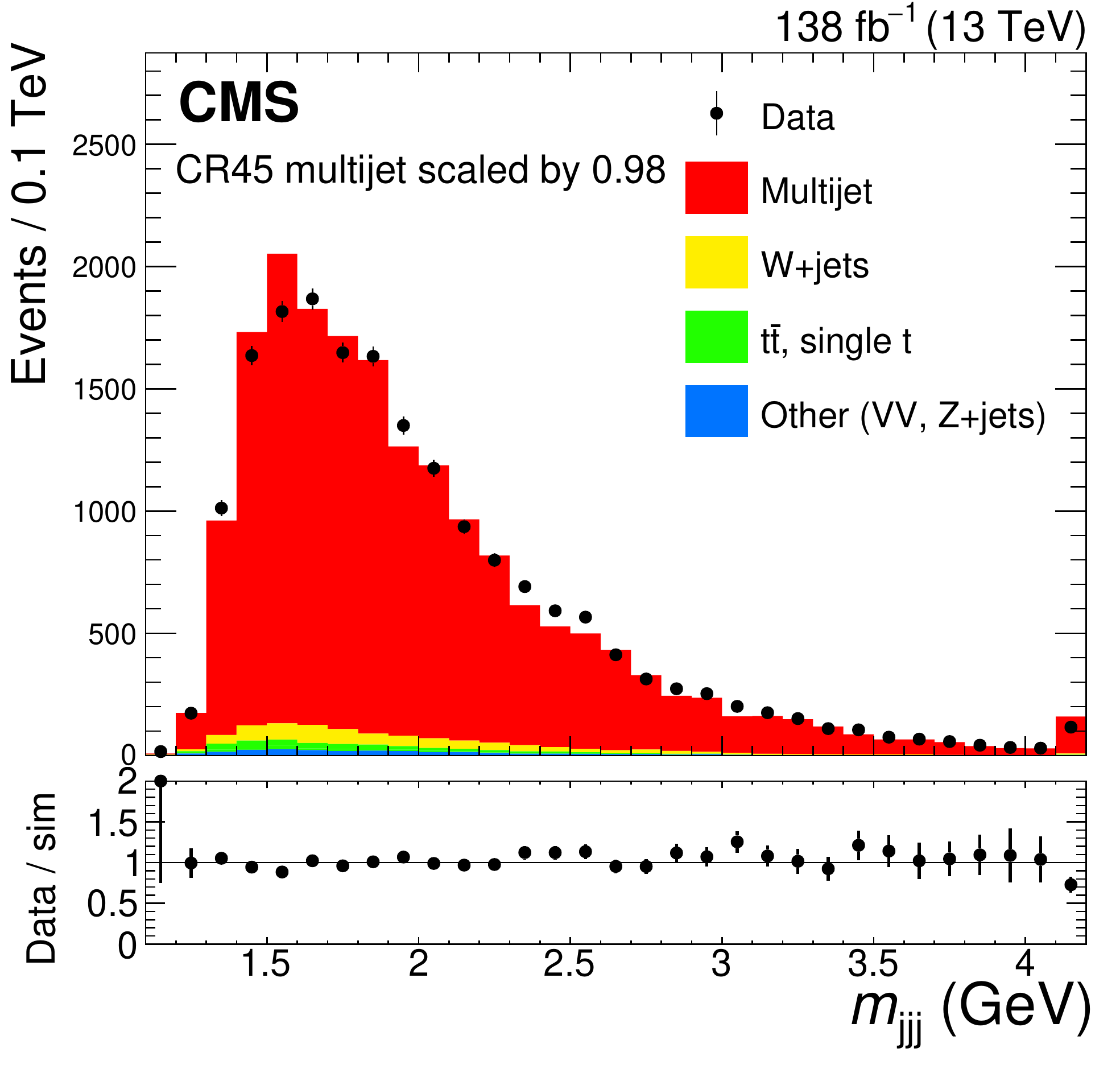}
\includegraphics[width=0.42\textwidth]{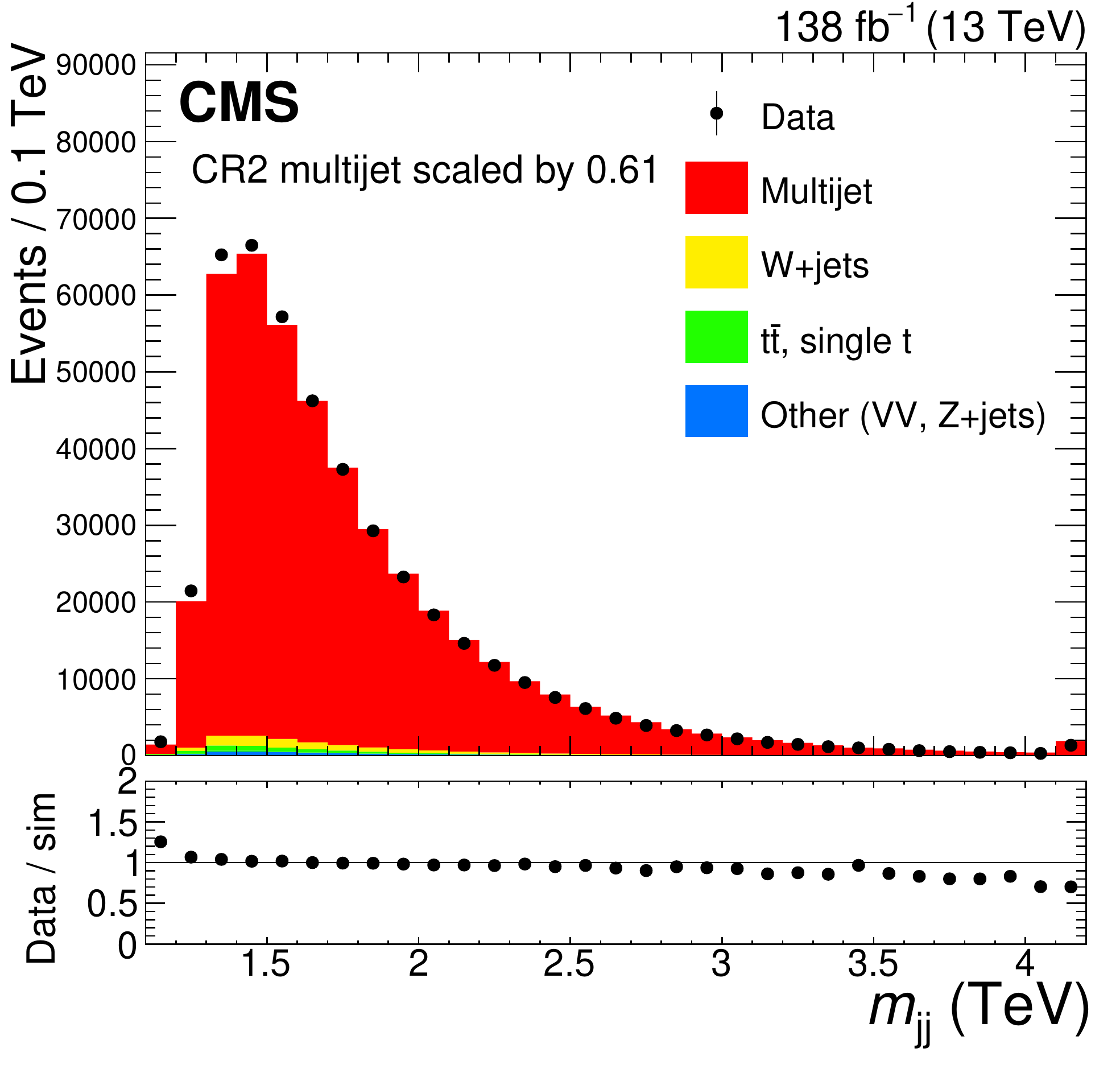}
\includegraphics[width=0.42\textwidth]{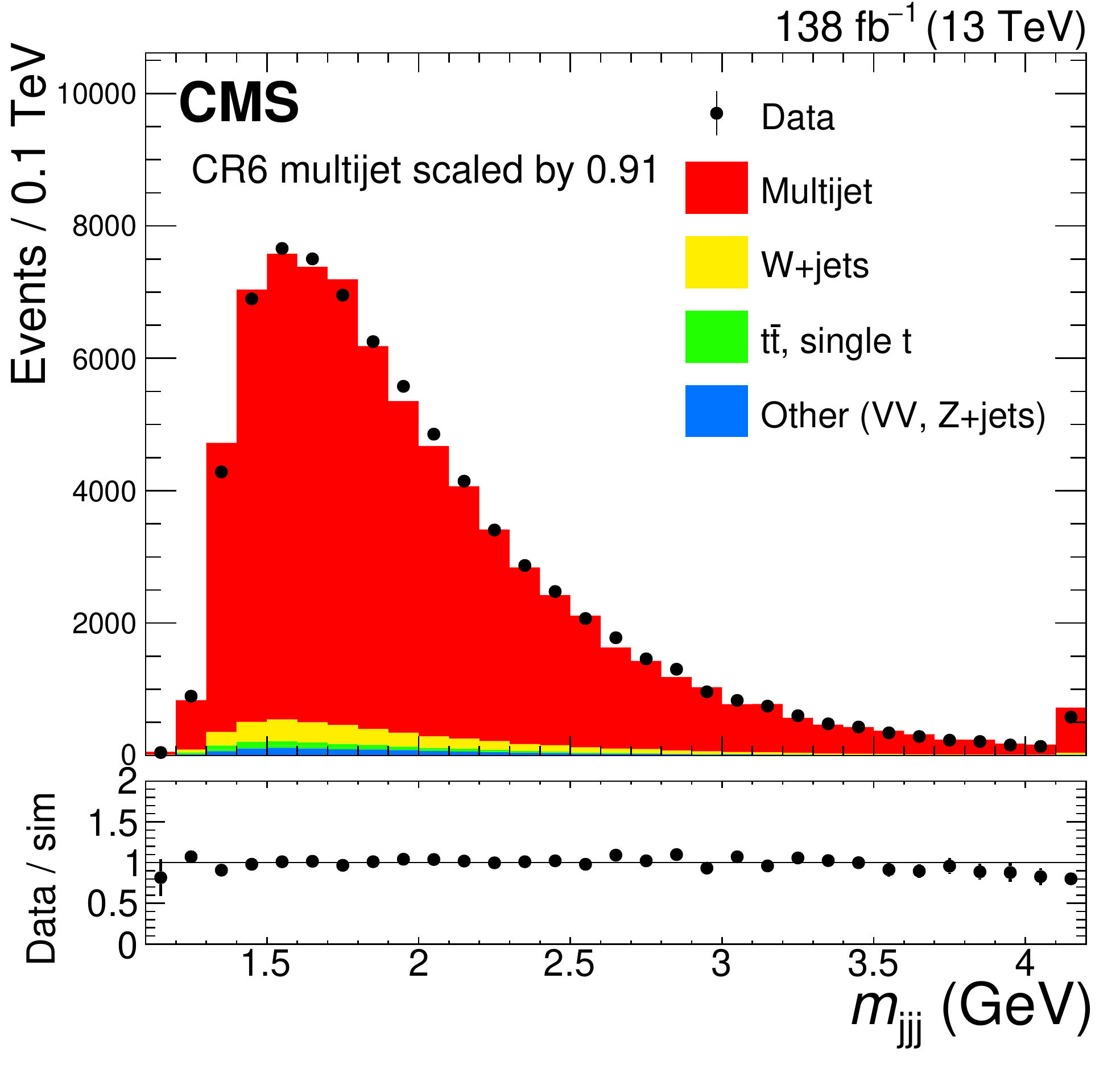}
\raggedright
\includegraphics[width=0.42\textwidth]{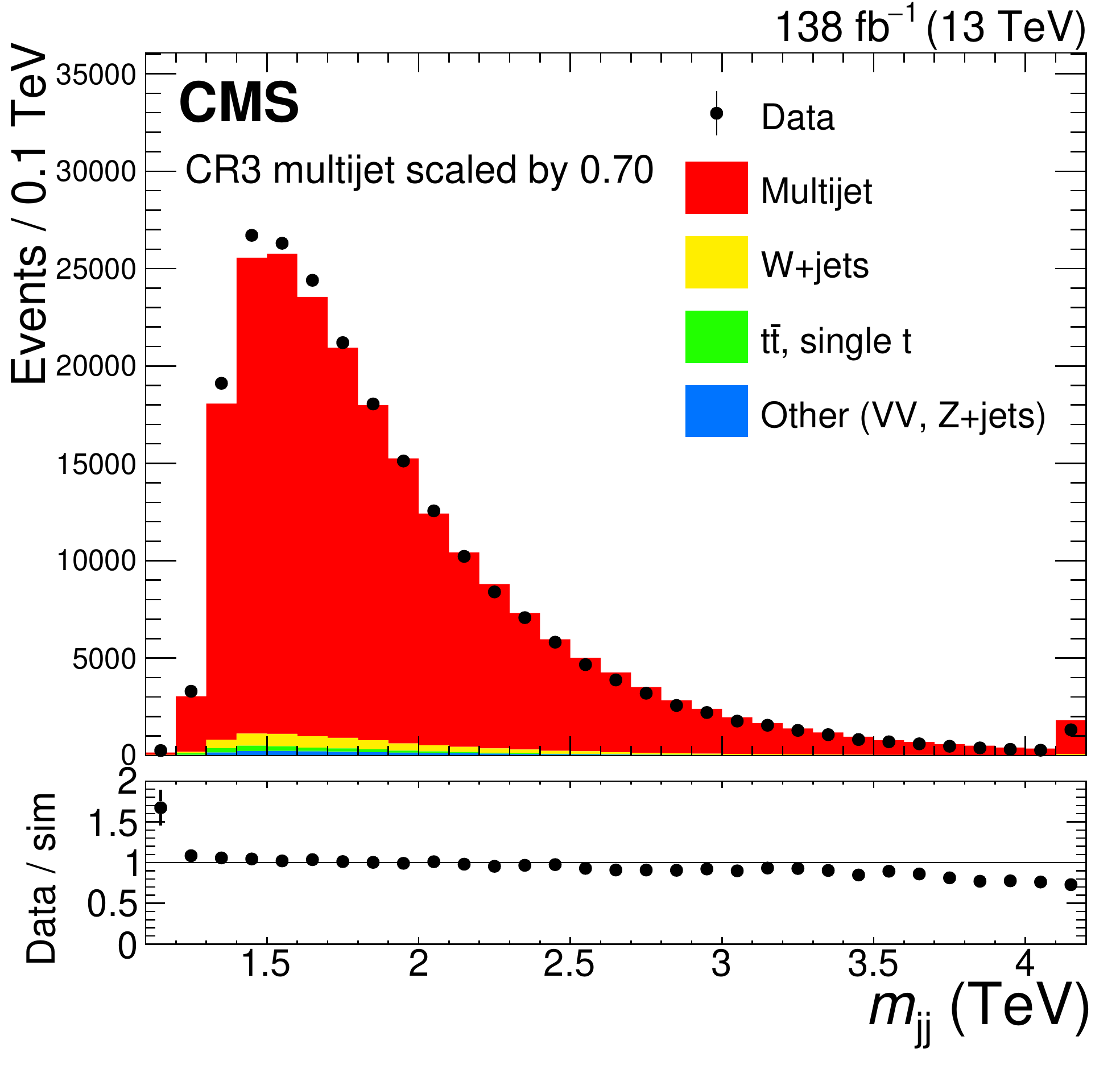}
\caption{
Invariant mass distributions of the reconstructed triboson systems for control regions in data (black markers) and simulated events (histograms).
The \Mjj distributions for CR1, CR2, CR3 are presented in the left column, upper to lower rows, respectively;
the \Mjjj distributions for control regions CR45 and CR6 are presented in the right column, upper and middle rows, respectively.
The simulation is corrected by SFs, and the QCD multijet background is scaled to the data yields.}
\label{fig:CRs_Observables}
\end{figure*}

The dominant background in all SRs consists of QCD multijet events, making up 60--80\% of the total.
As the \textsc{DeepAK8} tagger rejects the majority of these events, only a few of them remain in the SRs according to simulation.
Thus, we estimate this background contribution directly from the data using CRs.
The five CRs are defined by inverting at least one tagger condition, as described in Section~\ref{subsec:Calibration_of_q/g}. The selected jets in these regions possess similar kinematic properties to the ones in the corresponding SRs.
The \Mjj (\Mjjj) distributions in CRs 1--3 (4--6) are shown in Fig.~\ref{fig:CRs_Observables}, where the SF-corrected simulation is normalized to the data.
After subtracting the other background processes estimated from simulated samples from the data, the resulting \Mjj (\Mjjj) distributions are used to predict the shape of the QCD multijet background in the corresponding SRs.
This shape compatibility has been validated in simulation in multiple selections, and the \Mjj (\Mjjj) distributions agree within the statistical uncertainties over the entire spectra.
The a priori normalization in the SRs is taken from the SF-corrected simulation.
All other smaller background contributions such as \Wjets, \ttbar, single \PQt quark, diboson, $\ttbar\PV$, and triboson production are taken from simulation.

\section{Systematic uncertainties} \label{sec:uncertainties}

Systematic uncertainties are taken into account in the background estimation and the signal prediction.
For each source of uncertainty, a nuisance parameter is assigned, which is constrained by the data in the six SRs.
These are summarized in Table~\ref{tab:SysUnc}.

\begin{table*}[!htb]
\topcaption{Sources of systematic uncertainties accounted for in the analysis.
The first three sets of uncertainty sources originate from the tagger calibration.
It is also indicated whether the uncertainties are evaluated for background (B) and/or signal (S),
whether the uncertainty affects shape and/or rate, and the total number of nuisance parameters used per source.}
\label{tab:SysUnc}
\centering
\cmsTable{
\begin{scotch}{lcccc}
Sources                                                            & B or S & Effect on & Magnitude & Nuisance parameters     \\ \hline
Parton shower + selection bias for \PW, \Rlqq                      & B+S & Shape+rate &    \NA  &  4 for \dPWorWH $\times$ LL, LH         \\
Parton shower + selection bias for $\PQt^{2}$                      & B   & Shape+rate &    \NA  &  2(+4) for \dPWorWH LL, LH (LL, \ldots, HH) \\
Parton shower + selection bias for $\PQt^{3,4}$, $\PR^{3\PQq,4\PQq}$ & B+S & Shape+rate &  \NA  &  4 for \dPWorWH $\times$ HL, HH         \\
Parton shower + selection bias for \qg                             & B    & Shape+rate &   \NA  &  2(+4) for \dPWorWH LL, LH (LL, \ldots, HH) \\ [\cmsTabSkip]

Proxy uncertainty for \Rlqq                                        & S    & Rate        & 10--35\%&  2, for \dPWorWH  \\
Proxy uncertainty for $\PR^{3\Pq,4\Pq}$                            & S    & Rate        & 12--43\%&  2, for \dPWorWH  \\
Proxy uncertainty for unmatched                                    & S    & Rate        &   100\% &  2, for \dPWorWH  \\ [\cmsTabSkip]

High-\pt extrapolation for \PW                                     & S    & Rate        &   100\% &  2, for \dPWorWH  \\
High-\pt extrapolation for \Rlqq                                   & S    & Rate        & 23--30\%&  2, for \dPWorWH  \\
High-\pt extrapolation for \Rthreeq                                & S    & Rate        & 16--34\%&  2, for \dPWorWH  \\
High-\pt extrapolation for \Rfourq                                 & S    & Rate        & 24--33\%&  2, for \dPWorWH  \\ [\cmsTabSkip]
QCD multijet normalization                                         & B    & Rate        & 5--40\% &  5, common for SR4,5  \\
\ttbar normalization                                               & B    & Rate        &15--30\% &  5, common for SR4,5  \\
Other background normalization                                     & B    & Rate        & 30\%    &  5, common for SR4,5  \\  [\cmsTabSkip]
\MJJ, \MJJJ tail shape                                             & B    & Shape       & \NA     &  6, one for each SR  \\
\ttbar shape                                                       & B    & Shape       & \NA     &  6, one for each SR  \\
Pileup and integrated luminosity                                   & S    & Rate        & 1.7\%   &  1, common for all SRs \\
PDFs, renormalization and factorization scales                     & S    & Rate        & 1.4\%   &  1, common for all SRs \\
Jet energy scale and resolution                                    & S    & Shape       & \NA     &  2, common for all SRs \\
Jet mass scale                                                     & S    & Shape       & \NA     &  1, common for all SRs \\
\end{scotch}
}
\end{table*}

\subsection{Systematic uncertainties in the scale factor estimation}
\label{sec:uncertainties_for_SF}

Systematic uncertainties in the signal and background rate and shape arise from the \textsc{DeepAK8} SF derivation.
Two uncertainty sources common to signal and background jets are considered, and an additional two only for signal, which are described in Section~\ref{sec:uncertainties_for_Signal}.
The common uncertainties are from parton shower variations, and the SF dependence on the jet subsample selection, referred to as the ``selection bias'' uncertainty in Table~\ref{tab:SysUnc}.

The SFs are derived using three different \ttbar simulation samples: the nominal sample is generated using \POWHEG with \PYTHIA8, a second one using \POWHEG with \HERWIG 7~\cite{Bellm_2016}, and a third one using \MGvATNLO with \PYTHIA8.
The maximum difference of the three resulting SFs is symmetrized and assigned as the parton shower uncertainty for the \PW, $\PQt^2$, and $\PQt^{3,4}$ SFs.
For the \qg SFs, the parameters controlling the parton shower behavior in the QCD multijet \PYTHIA sample are varied to derive an uncertainty.
The resulting uncertainty bands are shown in Fig.~\ref{fig:SFs}, combined with the significantly smaller statistical uncertainty.

The bias in the SF calculation due to the selection conditions defining the jet subsample is estimated by performing closure tests in several validation regions such as jet mass sidebands.
The maximum nonclosure observed amounts to 10\% for \PW, $\PQt^{3,4}$, and \qg jets.
Because of the limited numbers of events in the validation regions for $\PQt^2$ jets and for jets not matching any of these categories, a 100\% uncertainty is assigned to those.
Uncertainties in the parton shower modeling and those arising from the selection bias are added in quadrature, and are assigned a single nuisance parameter for each matched jet in each LL, LH, HL, or HH bin.
The per-jet variation is treated as fully correlated.
Effects on both rate and shape of the \Mjj (\Mjjj) distributions are considered.
The overall rate uncertainties due to this variation amount to about 35\% for SRs 1--3 and SR6, 52\% for SR4, and 45\% for SR5.
These values are driven by the SF uncertainty on \qg jets, which constitute 75--90\% of the highest mass jets in the SRs.

\subsection{Systematic uncertainties in the background estimation}
\label{sec:uncertainties_for_BKG}

For the shape of the dominant QCD multijet background, we account for an additional uncertainty in the tail shape of the \Mjj (\Mjjj) distributions.
This uncertainty is derived in the CRs by comparing the QCD multijet prediction in simulation to the data.
A linear fit is performed to the ratio of the data and the simulation.
The resulting $\pm 2$ standard deviation bands are used as shape variations of the \Mjj (\Mjjj) distributions in the SRs.
A single nuisance parameter with a Gaussian prior is used for each SR.
This shape uncertainty allows the tails of the distributions to be adjusted by the data, accounting for effects that could lead to differences between CRs and SRs, \eg, a potential residual mass correlation of the taggers.

The uncertainty in the normalization of the QCD multijet background is taken as the normalization difference between data and SF-corrected simulation in the corresponding CRs.
These differences range from 9 to 40\% for SRs 1--3 and SR6, and 5\% for SR4--5.
For the top quark production rate, uncertainties in the normalization to NNLO and NLO predictions and missing higher orders are accounted for and are in the range 15--30\%. In addition, uncertainties in the \ttbar shape are derived by varying the top quark \pt spectrum based on the measurements in Refs.~\cite{CMS:2016oae,CMS:2019esx}.
For the other background processes, which are treated collectively, a 30\% normalization rate uncertainty is assigned for all SRs.
Because of their similarity, the same normalization nuisance parameters are used for SRs 4 and 5. All rate uncertainties are estimated using a log-normal prior.

\subsection{Signal systematic uncertainties}  \label{sec:uncertainties_for_Signal}

The integrated luminosities of the 2016, 2017, and 2018 data-taking periods are individually known with uncertainties in the 1.2--2.5\% range~\cite{Sirunyan:2021qkt,CMS-PAS-LUM-17-004,CMS-PAS-LUM-18-002}, while the total 2016--2018 integrated luminosity has an uncertainty of 1.6\%.
The simulated PU distribution is scaled to match data using an effective total inelastic cross section of 69.2\unit{mb}.
The uncertainty in this procedure is evaluated by varying the total inelastic cross section by $\pm$4.6\%~\cite{Sirunyan:2018nqx}.
This results in a 0.5\% uncertainty in the signal normalization in the SRs, which is combined with the integrated luminosity uncertainty for a total uncertainty of 1.7\%, implemented with a log-normal prior.

Renormalization $\mu_\text{R}$ and factorization $\mu_\text{F}$ scales and PDF uncertainties affecting the signal selection efficiency are evaluated per SR and mass point.
The scale uncertainties are obtained by varying $\mu_\text{R}$ and $\mu_\text{F}$ independently by factors of 1/2 and 2 (without considering the extreme cases of the opposite-direction variations).
The maximum value of these variations is taken as the prefit uncertainty.
For the overall scale uncertainty, a single nuisance parameter is used.
Its typical magnitude is up to 1.4\% for signal with $\MWKK\le4\TeV$.

The jet energy scale is varied by its uncertainty and the impact on the \Mjj (\Mjjj) distributions is taken to be the associated shape uncertainty.
Similarly, for the uncertainties in the jet energy and jet mass resolution, shape uncertainties are considered by varying the jets selected by the respective uncertainties.
All three uncertainty sources are implemented as nuisance parameters using Gaussian priors.

All of the above signal uncertainties only have a small impact on the final result.
The largest signal uncertainty originates from the \textsc{DeepAK8} tagger SF correction procedure.
Four different uncertainty sources are considered for the SFs applied to the signal jets.
The first two uncertainty sources are the parton shower and selection bias, and have only a small impact.
They are evaluated together with those of the background processes (described in Section~\ref{sec:uncertainties_for_SF}), using common nuisance parameters for signal and background jets.
Signal jets categorized as \PW, \Rlqq (\Rthreeq, \Rfourq) are assigned the same nuisance parameters as their corresponding proxy jets \PW ($\PQt^{3,4}$) and are constrained using the data in the SRs.
The other two sources of SF uncertainty, described below, are due to the differences between signal and proxy jets (proxy uncertainty),
and due to the significantly higher \pt that signal jets have compared to the SM jets (high-\pt extrapolation uncertainty).
Varying the SFs within these uncertainties has a major effect on the signal rates.

Although the signal jets share similar substructures with the corresponding SM proxies and also have similar \dPWorWH distributions, they are, with the exception of \PW boson jets, not the same objects.
For example, the flavor of the most energetic quarks might differ, the color flow structure might not be the same, and overall jet substructure kinematic properties could be different.
To account for all these differences, the shape difference of proxy and signal jets in six \dPWorWH bins above the 0.7 discriminant selection value (0.7--1.0 in 0.05 bins) is evaluated in simulation.
For each of these six bins, the relative difference between the proxy and the signal jets is taken as an uncertainty.
For signal jets categorized as $\PR^{3\Pq,4\Pq}$, for which the corresponding proxy jet category is $\PQt^{3,4}$, an additional uncertainty due to the difference observed between $\PQt^{3}$ and $\PQt^{4}$ is assigned.
It amounts to 5 and 10\% for the $\dPW\PH$ and \dPW discriminants, respectively.
The total resulting proxy uncertainties for \Rlqq, \Rthreeq, and \Rfourq signal jets lie in the ranges 10--35\%, 13--34\%, and 12--43\%, respectively.
This source of uncertainty has the largest effect on the rate for the merged signal.
Signal jets not matching any of these categories are assigned a 100\% proxy uncertainty.
The proxy uncertainty is evaluated separately for the \dPW (used in SR1) and $\dPW\PH$ (used in SR2--3) distributions, and is different for each signal mass scenario.

The high-\pt extrapolation uncertainty accounts for the fact that the SFs are derived in events containing jets with transverse momenta of a few hundred \GeVns, while the signal jets often have $\pt \geq 1\TeV$.
To account for this effect, the difference in the signal selection efficiency when using \HERWIG{}++~\cite{Bahr:2008pv} to perform the parton shower is evaluated with respect to the default \PYTHIA8 parton shower.
The uncertainty is evaluated separately for each of the four types of signal jets (\PW, \Rlqq, \Rthreeq, and \Rfourq) for the \dPW and $\dPW\PH$ discriminants.
It lies in the ranges 20--30 and 5--40\% for the merged and resolved signal, respectively.

The four \textsc{DeepAK8} tagger SF uncertainties (PS, selection bias, proxy, and high-\pt extrapolation) are considered as uncorrelated and result in a total uncertainty in the range 53--63 and 10--45\% for merged and resolved signals, respectively.

\section{Statistical analysis and results} \label{sec:Results}

\begin{figure*}[htbp]\centering
\includegraphics[width=0.41\textwidth]{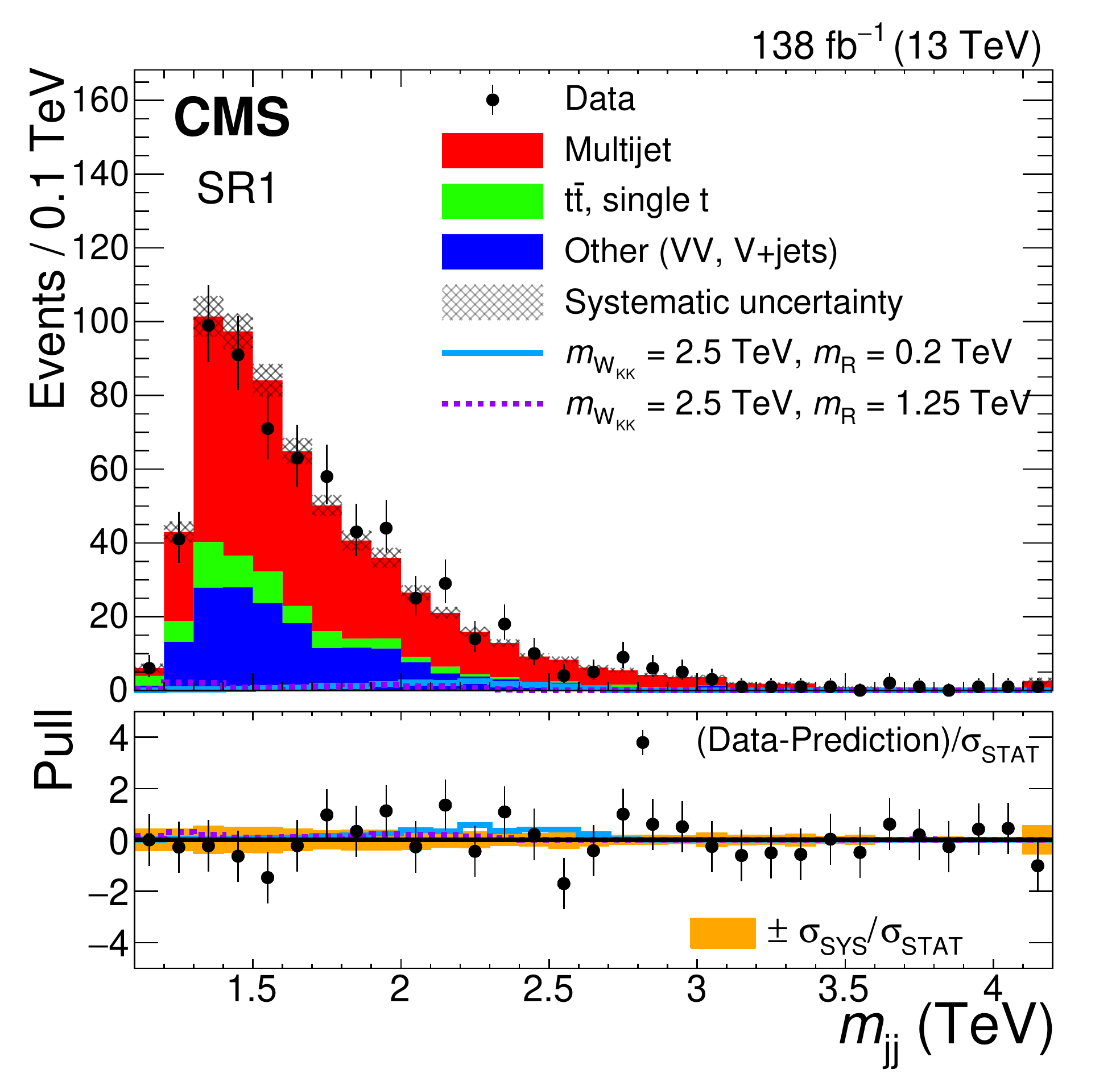}
\includegraphics[width=0.41\textwidth]{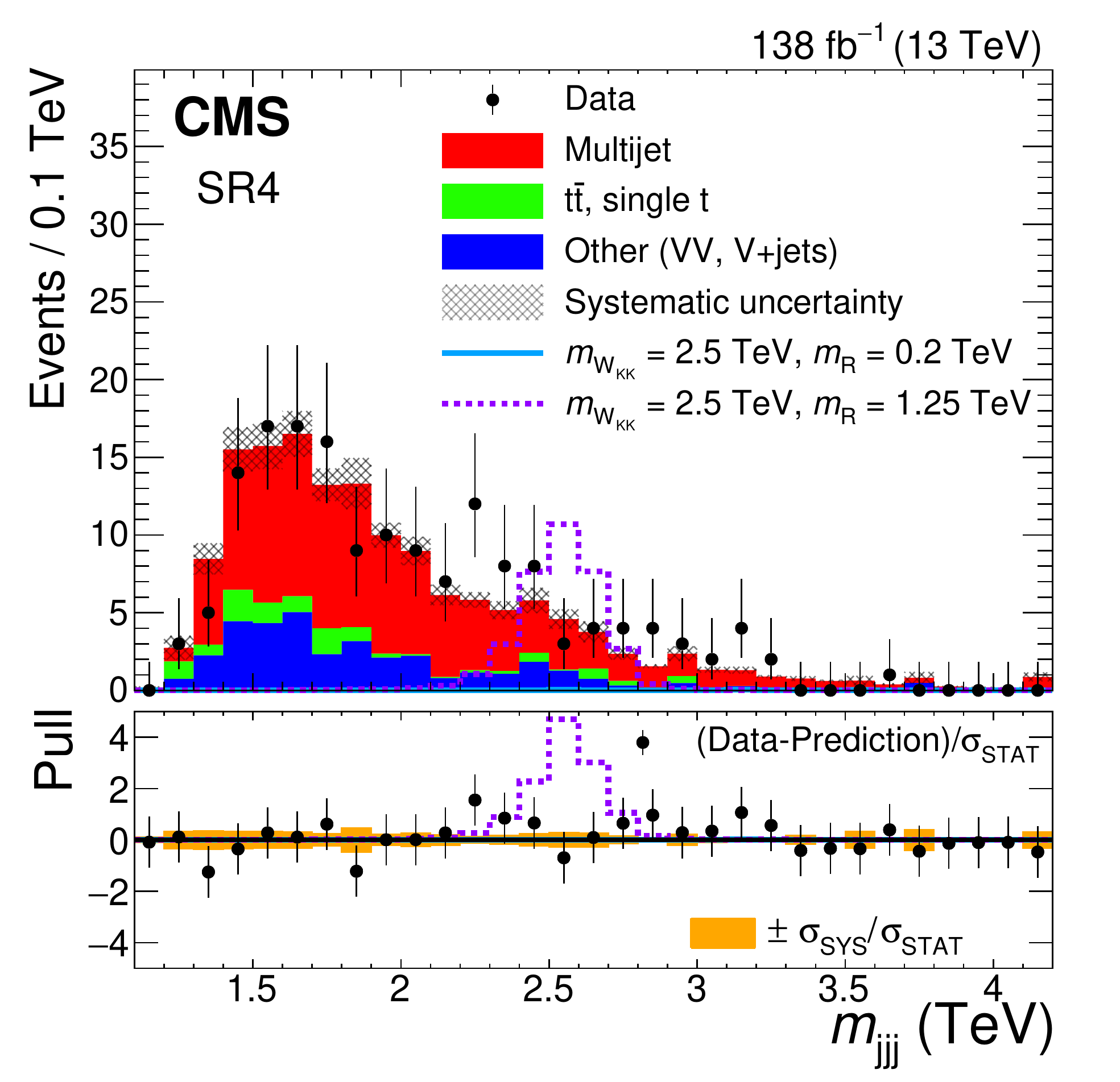}
\includegraphics[width=0.41\textwidth]{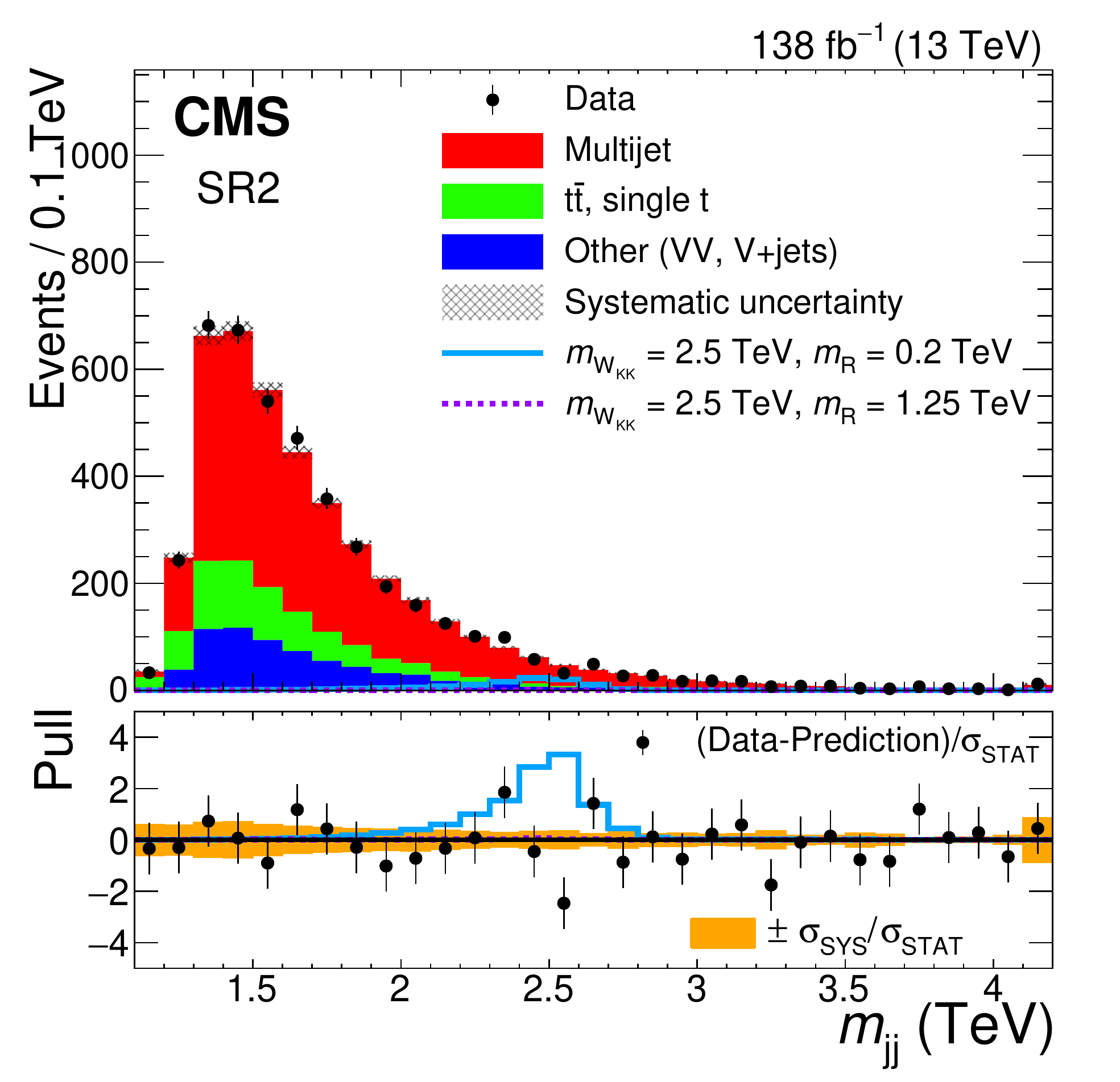}
\includegraphics[width=0.41\textwidth]{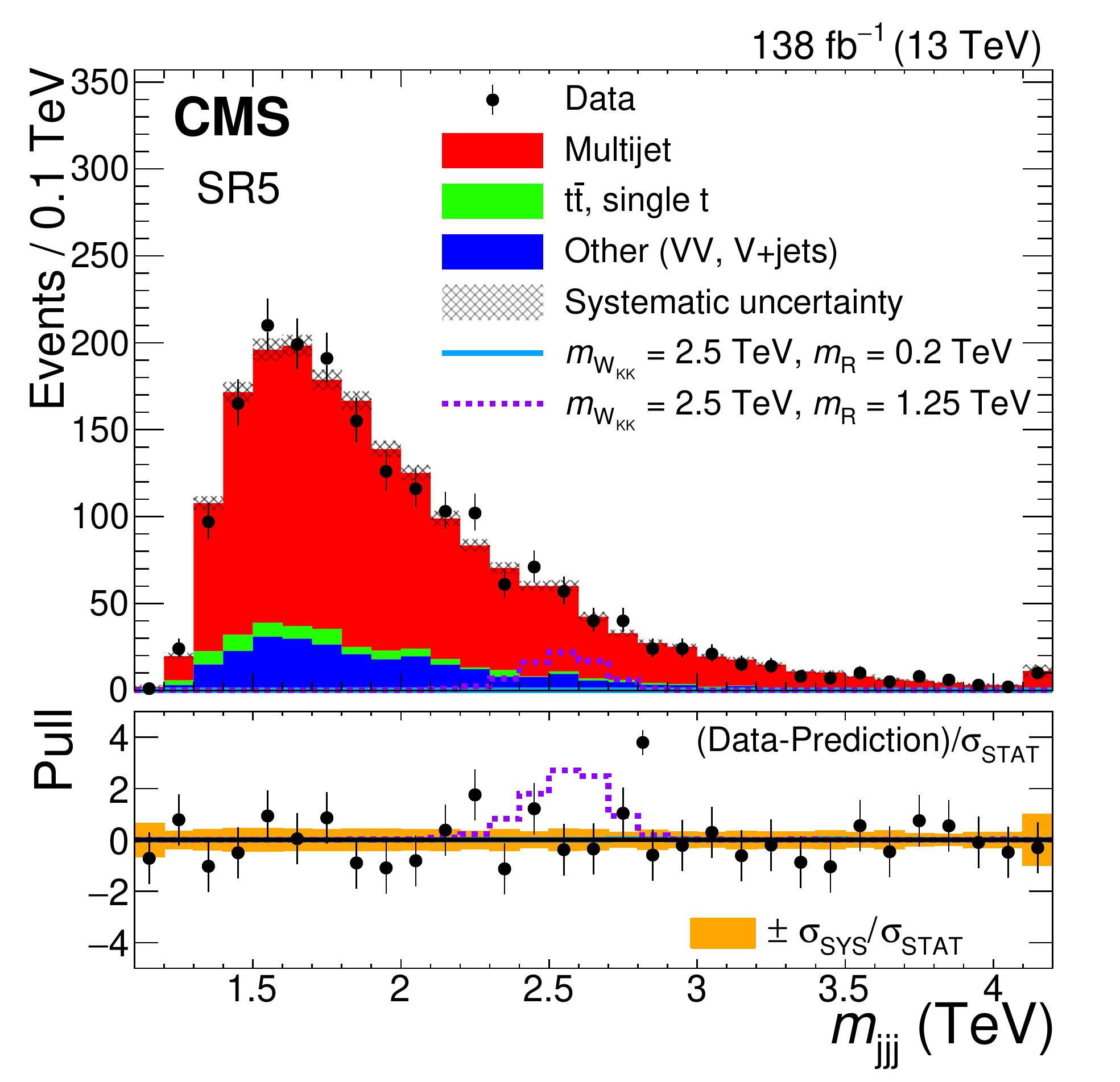}
\includegraphics[width=0.41\textwidth]{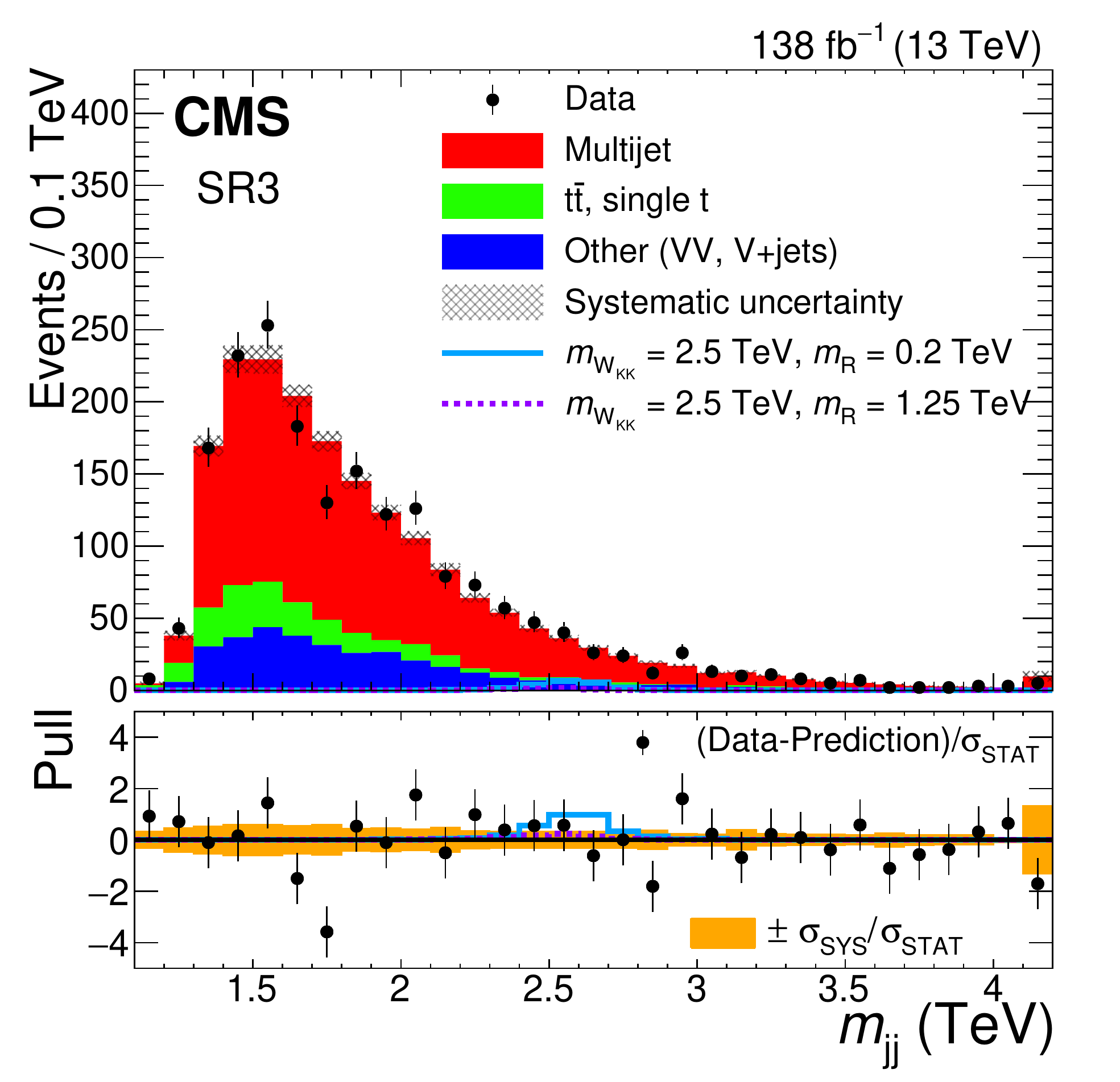}
\includegraphics[width=0.41\textwidth]{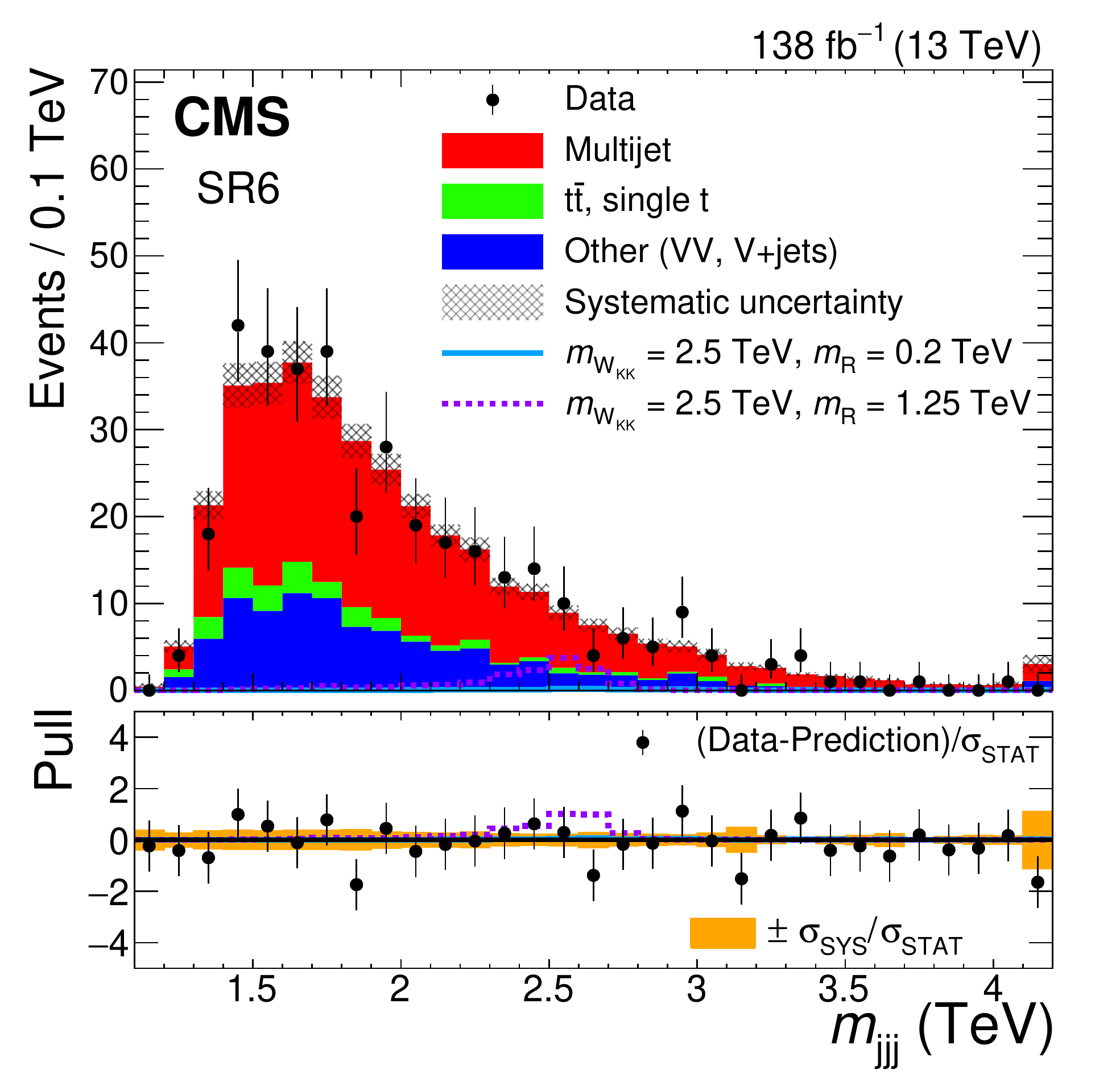}
\caption{
Post-fit distributions of the invariant mass of the reconstructed triboson system (\Mjj, \Mjjj) in data (black markers) and simulation (histograms)
for all SRs (SRs 1--3 in the left column and SRs 4--6 in the right column).
Systematic uncertainties are indicated by the shaded bands.
Signal examples are superimposed, normalized to the theoretical prediction for the production cross section of $\MWKK=2.5\TeV$ with $\MR=0.2\TeV$ (solid light blue line) and 1.25\TeV (dashed purple line).
The bottom panels show the pull distributions, indicating the difference between the data and background prediction, divided by the statistical uncertainty in the background,
with error bars representing the statistical uncertainty and shaded bands showing the one standard deviation systematic uncertainty, normalized by the statistical uncertainty.}
\label{fig:Postfit}
\end{figure*}

\begin{figure}[htb!]
\centering
\includegraphics[width=\cmsFigWidth]{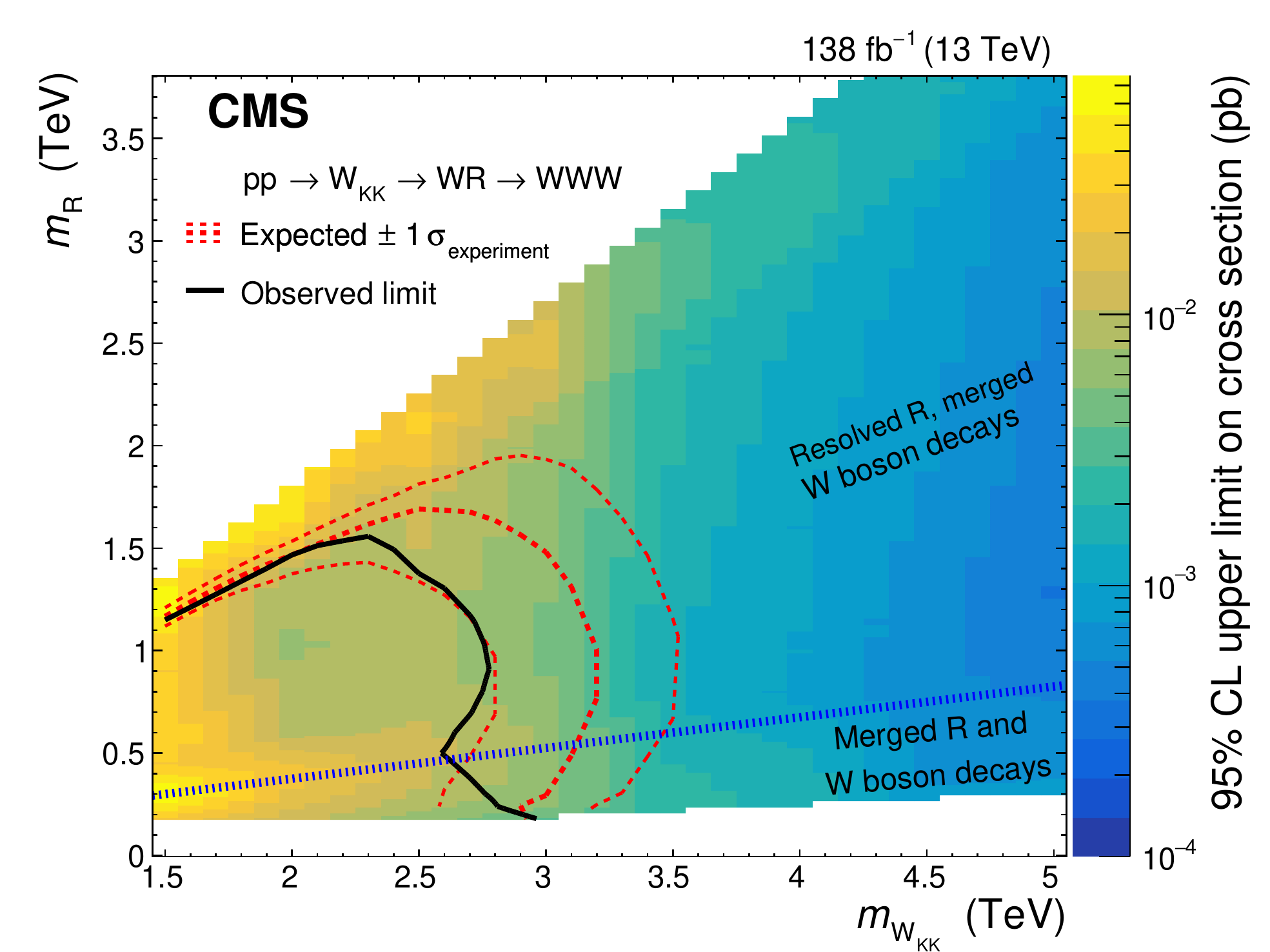}
\caption{
Expected (red dashed lines) and observed (solid black line) lower limits at 95\% \CL on the \WKK and \PR resonance masses for the particular parameters of the explored model.
The colored area indicates the observed upper limit on the product of the signal cross section and the branching fraction to three \PW bosons.
The blue dashed line indicates the border between the merged and resolved decay topologies probed.
A signal with \MR lower than 180\GeV is not considered in this search to maintain on-shell \PW bosons, while for $\MWKK > 3\TeV$, we only consider $\MR > 0.06\,\MWKK$.}
\label{fig:Limits3}
\end{figure}

The final \Mjj (\Mjjj) distributions for the SRs after performing a binned maximum likelihood fit in all six SRs simultaneously are shown in Fig.~\ref{fig:Postfit}.
No signal-like excess over the background expectation is observed in the data.
Upper limits at 95\% confidence level (\CL) are set on the
production cross section of a potential resonance signal as functions of the \WKK and \PR resonance masses.
The limits are set following the modified frequentist approach as described in Refs.~\cite{Junk:1999kv,Read:2002hq} and the definition of the profile likelihood test statistic as in Ref.~\cite{CMS-NOTE-2011-005} using an asymptotic approximation~\cite{Cowan:2010js}.
Figure~\ref{fig:Limits3} shows the limits on the product of the \WKK production cross section and the branching fraction to three \PW bosons.

We exclude \WKK resonances decaying in cascade via a scalar radion \PR to three \PW bosons at 95\% \CL with \MWKK up to 3\TeV for the lowest \MR of 200\GeV probed using the model provided in Refs.~\cite{Agashe:2016rle,Agashe:2016kfr,Agashe:2017wss,Agashe:2018leo}.
The highest \MR value excluded is 1.5\TeV for $\MWKK=2.3\TeV$.
The lower limits set on the production cross sections range from 70\unit{fb} at $\MWKK=1.5\TeV$ down to 0.5\unit{fb} at $\MWKK=5\TeV$.
The observed limits set in the \MWKK-\MR plane are weaker than the expected ones because of a mild excess of data events observed in SR4 around $\Mjjj = 3\pm 0.3\TeV$, which, however, exhibits no resonant structure.

For the resolved case, most of the sensitivity originates from SR4, complemented by SR5. For the merged case, SR2 and SR3 dominate the sensitivity and contribute roughly equally.
The SR1 and SR6 recover sensitivity to events where one \PW boson has relatively low \pt or mass.

\section{Summary}
\label{sec:sum}

A search for resonances decaying in cascade via a radion \PR to three \PW bosons, $\WKK \to \PW\PR \to \PW\PW\PW$, in the all-hadronic final state has been presented.
The search is performed in proton-proton collision data at a center-of-mass energy of 13\TeV, corresponding to a total integrated luminosity of 138\fbinv.
The final states include two or three massive, large-radius jets containing the decay products of the hadronically decaying \PW bosons.
The two-jet case corresponds to events where the radion decay products are reconstructed as a single merged jet.
The three-jet case corresponds to events where each \PW boson from the radion decay is reconstructed as a single merged jet.
In this analysis and the analysis in the single-lepton channel reported in Ref.~\cite{CMS-B2G-20-001}, previously unexplored signatures are probed, using novel jet substructure techniques.
In particular, a dedicated radion tagger based on a neural network, targeting simultaneously three different radion decay topologies, has been developed. This tagger has been calibrated with a novel ``matrix method''.
These techniques are also applicable to the identification of $\PH \to4\Pq$ and $\PH \to \Pq\Pq\Pell\PGn$ decays of Lorentz-boosted Higgs bosons.
Exclusion limits are set on the product of the production cross section and the branching fraction to three \PW bosons in an extended warped extra-dimensional model.
This result and the analysis in the single-lepton channel~\cite{CMS-B2G-20-001} are the first of their kind, and constrain the parameters of this model for the first time.

\begin{acknowledgments}
 We congratulate our colleagues in the CERN accelerator departments for the excellent performance of the LHC and thank the technical and administrative staffs at CERN and at other CMS institutes for their contributions to the success of the CMS effort. In addition, we gratefully acknowledge the computing centers and personnel of the Worldwide LHC Computing Grid and other centers for delivering so effectively the computing infrastructure essential to our analyses. Finally, we acknowledge the enduring support for the construction and operation of the LHC, the CMS detector, and the supporting computing infrastructure provided by the following funding agencies: BMBWF and FWF (Austria); FNRS and FWO (Belgium); CNPq, CAPES, FAPERJ, FAPERGS, and FAPESP (Brazil); MES and BNSF (Bulgaria); CERN; CAS, MoST, and NSFC (China); MINCIENCIAS (Colombia); MSES and CSF (Croatia); RIF (Cyprus); SENESCYT (Ecuador); MoER, ERC PUT and ERDF (Estonia); Academy of Finland, MEC, and HIP (Finland); CEA and CNRS/IN2P3 (France); BMBF, DFG, and HGF (Germany); GSRI (Greece); NKFIA (Hungary); DAE and DST (India); IPM (Iran); SFI (Ireland); INFN (Italy); MSIP and NRF (Republic of Korea); MES (Latvia); LAS (Lithuania); MOE and UM (Malaysia); BUAP, CINVESTAV, CONACYT, LNS, SEP, and UASLP-FAI (Mexico); MOS (Montenegro); MBIE (New Zealand); PAEC (Pakistan); MSHE and NSC (Poland); FCT (Portugal); JINR (Dubna); MON, RosAtom, RAS, RFBR, and NRC KI (Russia); MESTD (Serbia); MCIN/AEI and PCTI (Spain); MOSTR (Sri Lanka); Swiss Funding Agencies (Switzerland); MST (Taipei); ThEPCenter, IPST, STAR, and NSTDA (Thailand); TUBITAK and TAEK (Turkey); NASU (Ukraine); STFC (United Kingdom); DOE and NSF (USA).

\hyphenation{Rachada-pisek} Individuals have received support from the Marie-Curie program and the European Research Council and Horizon 2020 Grant, contract Nos.\ 675440, 724704, 752730, 758316, 765710, 824093, 884104, and COST Action CA16108 (European Union); the Leventis Foundation; the Alfred P.\ Sloan Foundation; the Alexander von Humboldt Foundation; the Belgian Federal Science Policy Office; the Fonds pour la Formation \`a la Recherche dans l'Industrie et dans l'Agriculture (FRIA-Belgium); the Agentschap voor Innovatie door Wetenschap en Technologie (IWT-Belgium); the F.R.S.-FNRS and FWO (Belgium) under the ``Excellence of Science -- EOS" -- be.h project n.\ 30820817; the Beijing Municipal Science \& Technology Commission, No. Z191100007219010; the Ministry of Education, Youth and Sports (MEYS) of the Czech Republic; the Deutsche Forschungsgemeinschaft (DFG), under Germany's Excellence Strategy -- EXC 2121 ``Quantum Universe" -- 390833306, and under project number 400140256 - GRK2497; the Lend\"ulet (``Momentum") Program and the J\'anos Bolyai Research Scholarship of the Hungarian Academy of Sciences, the New National Excellence Program \'UNKP, the NKFIA research grants 123842, 123959, 124845, 124850, 125105, 128713, 128786, and 129058 (Hungary); the Council of Science and Industrial Research, India; the Latvian Council of Science; the Ministry of Science and Higher Education and the National Science Center, contracts Opus 2014/15/B/ST2/03998 and 2015/19/B/ST2/02861 (Poland); the Funda\c{c}\~ao para a Ci\^encia e a Tecnologia, grant CEECIND/01334/2018 (Portugal); the National Priorities Research Program by Qatar National Research Fund; the Ministry of Science and Higher Education, projects no. 14.W03.31.0026 and no. FSWW-2020-0008, and the Russian Foundation for Basic Research, project No.19-42-703014 (Russia); MCIN/AEI/10.13039/501100011033, ERDF ``a way of making Europe", and the Programa Estatal de Fomento de la Investigaci{\'o}n Cient{\'i}fica y T{\'e}cnica de Excelencia Mar\'{\i}a de Maeztu, grant MDM-2017-0765 and Programa Severo Ochoa del Principado de Asturias (Spain); the Stavros Niarchos Foundation (Greece); the Rachadapisek Sompot Fund for Postdoctoral Fellowship, Chulalongkorn University and the Chulalongkorn Academic into Its 2nd Century Project Advancement Project (Thailand); the Kavli Foundation; the Nvidia Corporation; the SuperMicro Corporation; the Welch Foundation, contract C-1845; and the Weston Havens Foundation (USA).
\end{acknowledgments}

\bibliography{auto_generated}
\cleardoublepage \appendix\section{The CMS Collaboration \label{app:collab}}\begin{sloppypar}\hyphenpenalty=5000\widowpenalty=500\clubpenalty=5000\input{B2G-21-002-authorlist.tex}\end{sloppypar}
\end{document}

%% file: B2G-21-002-authorlist.tex
\cmsinstitute{Yerevan~Physics~Institute, Yerevan, Armenia}
A.~Tumasyan
\cmsinstitute{Institut~f\"{u}r~Hochenergiephysik, Vienna, Austria}
W.~Adam\cmsorcid{0000-0001-9099-4341}, J.W.~Andrejkovic, T.~Bergauer\cmsorcid{0000-0002-5786-0293}, S.~Chatterjee\cmsorcid{0000-0003-2660-0349}, K.~Damanakis, M.~Dragicevic\cmsorcid{0000-0003-1967-6783}, A.~Escalante~Del~Valle\cmsorcid{0000-0002-9702-6359}, R.~Fr\"{u}hwirth\cmsAuthorMark{1}, M.~Jeitler\cmsAuthorMark{1}\cmsorcid{0000-0002-5141-9560}, N.~Krammer, L.~Lechner\cmsorcid{0000-0002-3065-1141}, D.~Liko, I.~Mikulec, P.~Paulitsch, F.M.~Pitters, J.~Schieck\cmsAuthorMark{1}\cmsorcid{0000-0002-1058-8093}, R.~Sch\"{o}fbeck\cmsorcid{0000-0002-2332-8784}, D.~Schwarz, S.~Templ\cmsorcid{0000-0003-3137-5692}, W.~Waltenberger\cmsorcid{0000-0002-6215-7228}, C.-E.~Wulz\cmsAuthorMark{1}\cmsorcid{0000-0001-9226-5812}
% ...
\vskip\cmsinstskip
% ...
1:~Also at TU Wien, Wien, Austria\\
% ...
